\documentclass[aps,floatfix]{revtex4}

\usepackage{amssymb,hyperref,hypcap,
ae,tikz,color}
\usepackage{latexsym,textcomp,slashed,epsfig}
\usepackage[T1]{fontenc}
\usepackage{verbatim}

\usepackage{dcolumn}
\usepackage{bm}
\usepackage{amsmath}
\usepackage{graphicx}
\usepackage{listings}
\usepackage[T1]{fontenc}
\usepackage{xcolor}
\usepackage{lmodern}
\usepackage{listings}

\hypersetup{
	colorlinks=true,
	citecolor=blue,
    linkcolor=magenta,
    urlcolor=blue
	}

\begin{document}

\title{{\bf Propagation of Axial and Polar Gravitational Waves in Kantowski-Sachs Universe}}

\author{\bf{Sucheta Datta and Sarbari Guha}}

\affiliation{\bf Department of Physics, St.Xavier's College (Autonomous), Kolkata 700016, India}

\maketitle

\section*{Abstract}
In this paper, we apply the Regge-Wheeler formalism in our study of axial and polar gravitational waves in Kantowski-Sachs universe. The background field equations and the linearised perturbation equations for axial and polar modes are derived in presence of matter. To find the analytical solutions, we analyse the propagation of waves in vacuum spacetime. The background field equations in absence of matter are first solved by assuming that the expansion scalar $ \Theta $ to be proportional to the shear scalar $ \sigma $ (so that the metric coefficients are given by the relation $ a = b^n $, where $ n $ is an arbitrary constant). Using the method of separation of variables, the axial perturbation parameter $ h_0(t, r) $ is obtained from its wave equation. The other perturbation $ h_1(t, r) $ is then determined from $ h_0(t, r) $.
The anisotropy of the background spacetime is responsible for the damping of the axial waves. The polar perturbation equations are much more involved compared to their FLRW counterparts, as well as to the axial perturbations in Kantowski-Sachs background, and contain complicated couplings among the perturbation variables. In both the axial and polar cases, the radial and temporal solutions for the perturbations separate out as product. The temporal part of the polar perturbation solutions are plotted against time to obtain an order of magnitude estimate of the frequency of the propagating GWs, which is found to lie in the probable range of 1000-2000 Hz. Using standard observational data for the GW strain we have placed constraints on the parameters appearing in the polar perturbation solutions. The perturbation equations in presence of matter show that the axial waves can cause perturbations only in the azimuthal velocity of the fluid without deforming the matter field. But the polar waves must perturb the energy density, the pressure and also the non-azimuthal components of the fluid velocity. The propagation of axial and polar gravitational waves in Kantowski-Sachs and Bianchi I spacetimes is found to be more or less similar in nature.

\bigskip

KEYWORDS: Regge-Wheeler gauge, Kantowski-Sachs spacetime, Axial Gravitational waves, Polar Gravitational waves

\section{Introduction}

General Theory of Relativity (GR) is encapsulated in the Einstein field equations that relate the geometry of a spacetime with the corresponding matter distribution. As the Einstein equations are highly non-linear in nature, it is difficult to find exact solutions by any easy or straightforward method. Therefore, to find gravitational waves as solutions, one has to use approximation methods \cite{SAS}.

The theory of black hole perturbations was originally developed by Regge and Wheeler as a metric perturbation theory. While examining the stability of Schwarzschild singularity \cite{RW}, they incorporated small perturbations into the background space-time. By expressing the perturbations in terms of spherical harmonics, they proposed a standard gauge, now known as the Regge-Wheeler gauge, and derived a single Schrödinger-type differential equation for the perturbations. They found two gauge-independent linearised modes: the axial mode and the polar mode. The propagation equation for the polar waves derived in \cite{RW} was corrected by Zerilli \cite{ZER}. He also examined the problem of a particle falling into a Schwarzschild black hole \cite{ZERprd70}. The Regge-Wheeler (RW) procedure was adopted by Vishveshwara \cite{VISH} in Kruskal coordinates to examine the stability of the Schwarzschild exterior metric. The RW and Zerilli equations were re-derived using the variational approach in \cite{MON}. It was Chandrasekhar \cite{CHAND} who presented the metric perturbation theory in details. Evolution of stellar interior perturbations were matched with the exterior Schwarzschild perturbations at the stellar boundary in the papers by Seidel \textit{et al.} \cite{SEIDEL1}, which considered odd-parity perturbations of spherically symmetric stellar collapse models. Subsequently, Seidel \cite{SEIDEL2} developed a new set of gauge-invariant quantities from the RW variables while discussing the polar perturbations of spherical stellar collapse model. The equations were solved numerically by the May-White code. Fiziev \cite{FIZ} determined the exact solutions of the axial perturbations of the Schwarzschild metric in terms of the confluent Heun’s functions. The Laplace-transformed axial and polar perturbations for the RW equation in a de Sitter universe was examined in \cite{VIA}. More recently, Scharf \cite{SCHARF} has examined a non-standard cosmological model (homogeneous Datt or Kantowski-Sachs solution) in which visible matter and radiation are treated as perturbations on the vacuum solutions of the background Einstein equations. By using the mapping on the inner Schwarzschild solution and applying Zerilli’s perturbation theory, he simplified the linear perturbation equations and solved them by power series.

Malec and Wylezek \cite{M1} used the RW scheme to show that the axial modes in the FLRW spacetime obey Huygens principle in the radiation-dominated era, but in the matter-dominated universe, their propagation becomes wavelength-dependent. Kulczycki and Malec \cite{M2} studied the polar GWs in FLRW universe and found that these waves perturb both the density and non-azimuthal components of the velocity of the material medium, thereby contributing to the evolution of matter inhomogeneities and anisotropies. However, in the case of axial waves, the initial data can be chosen to decouple from matter, influencing only the azimuthal velocity, and cause local cosmological rotation. The Huygens principle is valid for both waves only in the radiative FLRW spacetimes. Although smooth axial waves do not interact with perfect fluids in radiative FLRW spacetimes \cite{M3}, but if the initial profiles are not smooth, then the wave pulses cause rotation of the radiation fluid and give rise to memory effect.
Rostworowski \cite{ROST1} has indicated that the RW formalism can be used as an alternative to study perturbations in a cosmological background, instead of the standard Bardeen's formalism of gauge-invariance \cite{BARDEEN}.
In \cite{ROST2}, he has shown that the general solution of the perturbation equations can be expressed in terms of two copies of a single master scalar that obeys scalar wave equation on the FLRW background (with a Regge-Wheeler / Zerilli type potential), and characterises the two gravitational degrees of freedom, with one scalar satisfying a transport type equation representing the matter perturbation.
The propagation of axial and polar waves using the RW gauge in FLRW spacetime has been discussed by Sharif and Siddiqa in the context of $ f(R,T) $ gravity \cite{SH1, SH2} and by Salti \textit{et al.} in Rastall gravity \cite{SALTI1, SALTI2}. In another paper \cite{SH3}, GWs produced by axially symmetric dissipative dust was examined in the context of $f(R)$ gravity. In a very recent paper \cite{SIDD}, Siddiqa \textit{et al.} have employed the RW axial perturbation scheme to solve the coupled system of differential equations of the perturbation parameters in the Starobinsky model of $ f(R) $ gravity, and have obtained analytical solutions for the radiation-dominated epoch and the de Sitter stage.

Axial and polar perturbations have been investigated in the gauge-invariant framework as well \cite{GMG, GS, CL1}. Gundlach and Martin-Garcia \cite{GMG} examined nonspherical perturbations of a spherically symmetric time-dependent background spacetime to study the production of gravitational waves using a $2+2$ reduction of the background spacetime. There exist three axial and seven polar gauge-invariant matter perturbations. For $l=0$ there are no axial perturbations, and for $l=1$, there exists a constraint on the scalar wave equation. They identified a set of true perturbation degrees of freedom admitting free initial data for the axial and for the $l\geqslant2$ polar perturbations. However, the gauge-invariant treatment breaks down for the polar $l=0$ and $l=1$ perturbations, and moreover, these perturbations do not couple to GWs. It may be noted that the equations (2.40) and (2.41) in \cite{SEIDEL2} were specific to the case of $l=2$ angular dependence. The metric perturbations of the Schwarzschild spacetime was studied using a covariant and gauge-invariant formalism by Martel and Poisson \cite{MAR}. Clarkson \textit{et al.} \cite{CL1} derived a complete system of master equations for the dust-dominated LTB spacetimes, and showed that unlike in FLRW cosmology where any perturbation can be split into independently-evolving scalar, vector and tensor (SVT) modes, such a split is not possible in the LTB model, where the modes get coupled. However, depending on the nature of their transformation on the surfaces of spherical symmetry, the perturbations can be decoupled into two independent modes, namely polar (or even) and axial (or odd). These modes are analogous to scalar and vector modes in the FLRW model, although further decomposition into tensor modes is not possible because non-trivial symmetric, transverse and trace-free rank-2 tensors cannot exist on $ S^2 $. The paper \cite{CL2} derives the numerical solution to the master equation of polar waves in LTB dust model.

It is well-known that exact solutions for homogeneous spacetimes in GR belongs to either Bianchi types or the Kantowski-Sachs model \cite{AKRC}. Kantowski-Sachs (KS) spacetime \cite{KS} is the only anisotropic but spatially homogeneous cosmology that does not come under the Bianchi classification \cite{BD1, BR1, KATORE}. Much before it became widely known with the publication by Kantowski and Sachs \cite{KS}, Datt \cite{DATT} was the first to discuss a generalisation of this type of metric \cite{SCHARF, KRAS}. These cosmological models exhibit spherical symmetry along with a translational symmetry and can be treated as non-empty analogs of a part of the extended Schwarzschild manifold \cite{COL77}. The effect of anisotropy observed in KS models is comparable with the FRLW class of models \cite{KATORE}. Of the LRS cosmologies of class II, the KS metric possesses positive curvature (2D scalar curvature), while the zero and negative curvature ones are respectively Bianchi I/ VII$ _0 $ (including flat Friedmann universes), and the Bianchi III model \cite{BR1, KATORE}. The general form of the line element can be expressed as \cite{SHAMIR1006} :
\begin{equation}
ds^2 = dt^2 -a^2(t)dr^2 -b^2(t)[d\theta ^2 + f_{\mathcal{K}}(\theta) d\phi ^2],
\end{equation}
where $ \mathcal{K} $ is the spatial curvature index of the universe.
For $ \mathcal{K} =0 $, $ f_{\mathcal{K}}(\theta) =\theta ^2 $, and the universe is Bianchi type I.
For $ \mathcal{K} =1 $, $ f_{\mathcal{K}}(\theta) =\sin ^2 \theta $, the model is Kantowski-Sachs and closed.
For $ \mathcal{K} =-1 $, $ f_{\mathcal{K}}(\theta) =\sinh ^2 \theta $, the model represents Bianchi III space-time, which is semi-closed.

Kantowski-Sachs cosmology has been studied by several authors \cite{WB1, WB2, GRON, MAT, BD2, KRORI, DABROWSKI}. Kantowski-Sachs string cosmology was discussed in \cite{BD1}. Dust Kantowski-Sachs models were studied in \cite{BD2} in the context of cyclic closed universes. Shamir \cite{SHAMIR1006} has studied the vacuum Kantowski–Sachs and Bianchi type III spacetimes in the $ f(R) $ theory of gravity. The paper \cite{SHABBIR} deals with the conformal vector fields of Kantowski–Sachs and Bianchi III spacetimes filled with perfect fluid in the $ f(R) $ theory of gravity. The paper \cite{KATORE} conducted a study of these two models in the $ f(R, T) $ theory. In \cite{SHAMIR1006}, and \cite{SHABBIR}, the vacuum field equations are solved by assuming the expansion scalar $ \Theta $ to be proportional to the shear scalar $ \sigma $. The article \cite{FAG} treats a perturbed Kantowski-Sachs model using numerical integration. In \cite{TER}, the authors derived exact solution for the GR field equations during the gravitational collapse in an anisotropic spacetime (generalization of KS model) with arbitrary curvature. They investigated the singularity formation and determined the null directions for GW propagation.

Using the framework of 1+1+2 covariant decomposition of spacetime, Keresztes \textit{et al.} \cite{BR1} studied the perfect-fluid perturbations of the Kantowski-Sachs universe with vanishing vorticity and a positive cosmological constant. The dynamics is described by six evolution equations for six harmonic coefficients that represent gravito-magnetic, kinematic and matter perturbations. 
In the geometrical optics limit, the coupled gravito-magnetic differential equations decouple and the shear and matter-gradient perturbations show wavelike evolutions to the leading order. The even and odd tensorial perturbations of the Weyl tensor behave as GWs travelling at the speed of light in vacuum. The modes denoting the shear and the matter-density gradient propagate at the speed of sound, while remaining out of phase by $ \pi /2 $. At the next higher order, GWs suffer damping, while the shear and matter waves undergo forced oscillations. Beyond the geometrical optics approximation, they have indicated the existence of direction-dependent dispersion relations. This treatment has also been extended to LRS class-II cosmologies \cite{BR2}.

In our earlier paper \cite{GD} we studied the propagation of axial gravitational waves in the anisotropic but spatially homogeneous Bianchi I universe, putting to use the Regge-Wheeler gauge. We found that the solutions of the vacuum perturbation equations yield the complete expressions for the axially perturbing variables in the form of product of four factors, each a function of one coordinate only. In the present paper, we adopt the RW gauge to investigate axial as well as polar waves in Kantowski-Sachs spacetime. Similar studies incorporating the RW perturbation scheme have been undertaken with FLRW metric as the background \cite{M1, M2, M3, SH1, SH2, ROST2, SALTI1, SALTI2}. Perturbations on KS model have been discussed in some other papers \cite{FAG, TER, BR1}, but the RW gauge has not been employed before. Our paper is organised as follows: Sec. II contains the background metric and the corresponding Einstein equations with and without matter distribution. The Regge-Wheeler gauge for axial and polar modes is introduced in Sec. III. Sec. IV carries a discussion on the energy-momentum tensor in presence of perturbations. In Sec. V, the linearised field equations are derived for both axial and polar perturbations, using suitable constraints in order to make our analysis tractable. We solve these perturbation equations analytically in the vacuum spacetime to determine the axial and the polar modes in Sec. VI. In Sec. VII we examine some sample plots of the solutions obtained in the case of polar waves. Subsequently we have shown how we can utilize the standard observational data to constrain the parameters appearing in  the perturbation solutions. We have considered the method discussed in \cite{SCHUTZ} to examine the effect of the waves on free particles, and have adapted it to the case of the waves propagating in a Kantowski Sachs background. We conclude with an analysis of our results and some remarks in Sec. VIII. In Appendix A, we present the field equations for polar perturbations by removing the constraint $\eta = 0 $ on the perturbation variable. We find that the equations for polar perturbations with $\eta \neq 0 $ are very much complicated. Hence we have assumed the background matter to be in the form of stiff perfect fluid in order to simplify the perturbation equations for the $\eta \neq 0 $ case. We know that during the period of high densities after the cosmic explosion, the matter content of the universe could have been in the form of a stiff fluid \cite{Zeldovich, Barrow, Chavanis}. It is also known that the study of gravitational waves which propagate in dust or in fluids with a realistic equation of state is quite problematic, whereas exact radiative solutions could be obtained when studying the propagation of gravitational waves in a stiff fluid \cite{Griffiths1, Griffiths2, BG1, FG, AG, BG2}, and these solutions may be regarded as the perturbations of the FRW models with a stiff fluid. In \cite{BD1} it was shown that a time-dependent axion field in a Kantowski-Sacks background behaves effectively as a stiff perfect fluid which is distributed homogeneously over space. In \cite{DABROWSKI}, the behaviour of growing-entropy stiff-fluid Kantowski-Sachs universes was investigated.

Throughout this paper, an overdot represents a derivative w.r.t $ t $ and a prime denotes a derivative w.r.t. $ r $, and geometrized units, i.e., $ 8\pi G = c = 1 $ are assumed.

\section{The unperturbed background metric and field equations}

The Kantowski-Sachs spacetime is defined in spherical polar coordinates by the line element:
\begin{equation} \label{1}
ds^2= dt^2- a^2(t) dr^2 -b^2(t) d\theta ^2 -b^2 (t)\sin^2 \theta d\phi ^2 ,
\end{equation}
where $ a(t) $ and $ b(t) $ are the scale factors for expansion parallel and perpendicular to the radial direction respectively.

Let us assume the background spacetime to be filled with a perfect fluid. If $u^\alpha$ be the fluid four-velocity, $ \rho $ be the energy density and $ p $ be the pressure, then the the energy-momentum tensor is given by
\begin{equation} \label{1a}
T_{\mu\nu}= (\rho+p) u_{\mu}u_{\nu} -p g_{\mu\nu}.
\end{equation}
We use the subscript `0' to indicate the energy density and pressure of the fluid for the background metric \eqref{1}. The corresponding field equations are as follows:
\begin{equation}\label{2a}
\frac{2 \dot{a}\dot{b}}{ab} + \frac{\dot{b}^2}{b^2} + \frac{1}{b^2} =\rho_0,
\end{equation}
\begin{equation}\label{2b}
-a^2\left( \frac{2\ddot{b}}{b} +\frac{{\dot{b}}^2}{b^2} + \frac{1}{b^2} \right) =a^2 p_0,
\end{equation}
\begin{equation}\label{2c}
-b^2\left( \frac{\ddot{a}}{a} +\frac{\ddot{b}}{b} +\frac{\dot{a} \dot{b}}{ab}\right) =b^2 p_0.
\end{equation}

The volume expansion $ \Theta $ and shear scalar $ \sigma $ for the metric \eqref{1} are given by
\begin{equation}\label{4a}
\Theta = \frac{\dot{a}}{a} + \frac{2\dot{b}}{b},  \hspace{0.5cm} \text{and} \hspace{0.5cm}  \sigma ^2 = \frac{1}{3}\left( \frac{\dot{a}}{a} - \frac{\dot{b}}{b} \right)^2 .
\end{equation}
If the ratio of the shear $ \sigma $ to the expansion $ \Theta $ is constant, the corresponding cosmological model remains anisotropic throughout its evolution \cite{GRON2, ROY1, ROY2, BAG, BALI}. In such cases, the expansion scalar is proportional to the shear scalar, and it is assumed that
\begin{equation}\label{4b}
a=b^n,
\end{equation}
where $ n $ is an arbitrary real number and $ n \neq 0, 1 $ for non-trivial solutions. We will use this relation to solve the vacuum field equations. 
Substituting \eqref{4b} in \eqref{4a}, we find that
\begin{equation}
\frac{\sigma ^2}{\Theta ^2} =\frac{1}{3}\left( \frac{n-1}{n+2} \right) ^2,
\end{equation}
which is constant for a given $ n $. Exact spatially homogeneous cosmologies with this ratio as constant, were studied in Ref. \cite{COL80}.

The continuity equation for the KS metric \eqref{1} is represented by:
\begin{equation} \label{E1}
\dot{\rho} + \left( \frac{\dot{a}}{a} + \frac{2\dot{b}}{b}\right) (\rho_0 +p_0) =0.
\end{equation}

In vacuum, the background field equations \eqref{2a}-\eqref{2c} are reduced to :
\begin{equation}\label{3a}
\frac{2 \dot{a}\dot{b}}{ab} +\frac{\dot{b}^2}{b^2} +\frac{1}{b^2} =0,
\end{equation}
\begin{equation}\label{3b}
\frac{2\ddot{b}}{b} +\frac{{\dot{b}}^2}{b^2} +\frac{1}{b^2} =0,
\end{equation}
\begin{equation}\label{3c}
\frac{\ddot{a}}{a} +\frac{\ddot{b}}{b} + \frac{\dot{a} \dot{b}}{ab} =0.  
\end{equation}
Combining \eqref{3a}, \eqref{3b} and \eqref{3c} we finally obtain
\begin{equation}\label{5c}
\frac{\ddot{a}}{a} + \frac{2\dot{a} \dot{b}}{ab} =0.
\end{equation}
Inserting equation \eqref{4b} in equation \eqref{5c}, we get  
\begin{equation}\label{5e}
\frac{\ddot{b}}{b} +(n+1)\frac{\dot{b}^2}{b^2} =0,
\end{equation}
which has the solution
\begin{equation}\label{5f}
b=[(n+2)(k_1 t +k_2)]^{\frac{1}{n+2}},
\end{equation}
where $ k_1 $ and $ k_2 $ are integration constants. Without loss of generality, $ k_2 $ can be set to zero, and we have
\begin{equation}\label{5fi}
b=[(n+2)(k_1 t)]^{\frac{1}{n+2}}.
\end{equation}

Using equations \eqref{4b} and \eqref{5fi} in equation \eqref{3c}, we get
\begin{equation}\label{5g}
n^2 \frac{\dot{b}^2}{b^2} + (n+1)\frac{\ddot{b}}{b} =0,
\end{equation}
which means that
\begin{equation}\label{5gi}
(n+2)^{(\frac{2}{n+2}-1)} k_1^2 (k_1t)^{(\frac{2}{n+2}-2)} \left[ \frac{n^2}{n+2} + (n+1)\left(\frac{1}{n+2}-1\right)  \right] =0.
\end{equation}
The term in the square brackets in \eqref{5gi} must vanish, so that
\begin{equation}\label{5h}
n^2+(n+1)(1-n-2) =0  \hspace{0.5cm} \Rightarrow \hspace{0.5cm} n=-1/2.
\end{equation}
Thus, we have
\begin{equation}\label{5hi}
b=\left(\frac{3}{2} k_1t\right) ^{2/3} =Kt^{2/3}, \hspace{0.5cm} \text{and} \hspace{0.5cm} a=b^{-1/2}.
\end{equation}
Here $ K=(3k_1/2)^{(2/3)} $ is a constant. We will use \eqref{5hi} in the calculations that follow.

\section{The perturbed metric and the Regge-Wheeler scheme}

We now incorporate small perturbations on the Kantowski-Sachs background. The perturbed metric can be written as
\begin{equation}\label{1b}
g_{\mu\nu} = g_{\mu\nu}^{(0)} + e h_{\mu\nu} + \mathcal{O}(e^2),
\end{equation}
where $ g_{\mu\nu}^{(0)} $ is the background metric \eqref{1} and $ h_{\mu\nu} $ denote the perturbations representing gravitational waves on it. Here $ e $ is a small parameter which gives a measure of the strength of perturbations, and $ \mathcal{O}(e^2) $ indicate the terms involving $ e^2 $ and other higher orders of $ e $, which are neglected in the subsequent calculations.

Throughout this paper we will follow the approach that we adopted in our previous paper \cite{GD}, in order to apply the Regge-Wheeler perturbation procedure \cite{RW} to study the GWs in Kantowski-Sachs universe.
In this scheme, under a rotation of the frame about the origin, the components $ h_{00} $, $ h_{01} $, $ h_{11} $ of the perturbation matrix transform like scalars, $ (h_{02} $, $ h_{03} $) and ($ h_{12} $, $ h_{13} $) transform like vectors, and $ h_{22} $, $ h_{23} $ and $ h_{33} $ transform like a second-order tensor. These scalars, vectors and tensors were expressed in terms of spherical harmonics $ Y_{lm} $ where $ l $ is the angular momentum and $ m $ is its projection on the $ z $-axis.
Since all values of $ m $ yield the same radial equation, the value $ m=0 $ was chosen, as a result of which the $ \phi $-dependence disappeared \cite{RW}. Even after these simplifications, the axial waves were given by three unknown functions of $ r $, and the polar waves contained seven unknown functions. The Regge-Wheeler gauge was introduced to find the canonical form of the axial and polar waves. The $ t $ and $ r $-solutions separated out as product in the final expressions.

In case of the odd (or axial) waves, there are only two non-zero components of $ h_{\mu\nu} $ \cite{M2} denoted by:
\begin{equation}\label{6a}
h_{t \phi} =h_0(t,r) \sin \theta (\partial_{\theta}Y) \hspace{0.2cm} \text {and} \hspace{0.2cm}
h_{r \phi} =h_1(t,r) \sin \theta (\partial_{\theta}Y).
\end{equation}
Thus, for the background metric \eqref{1}, the axially perturbed line element is given by:
\begin{equation}\label{6b}
ds^2= dt^2- a^2(t) dr^2 -b^2(t) d\theta ^2 -b^2(t) \sin ^2 \theta d\phi ^2
+2eh_0(t,r) \sin \theta (\partial_{\theta}Y) dtd\phi +2eh_1(t,r) \sin \theta (\partial_{\theta}Y) drd\phi + \mathcal{O}(e^2).
\end{equation}
Here, the spherical harmonics $ Y_{lm}(\theta, \phi) $ are denoted by $ Y $, where one chooses $m=0$. For wavelike solutions, $ l \geq 2 $ \cite{SH1}. Further $ Y_{lm}(\theta, \phi) $ satisfies the relation:
\begin{equation}\label{sph}
\partial_{\theta} \partial_{\theta} Y = -l(l+1)Y - \cot\theta (\partial_{\theta} Y).
\end{equation}

On the other hand, $ h_{\mu\nu} $ has three non-zero components for even (or polar) waves in the Regge-Wheeler gauge \cite{RW}. Following the Gerlach-Sengupta \cite{GS} formalism further developed by Gundlach and Martin-Garcia \cite{GMG}, Clarkson and others \cite{CL1} have expressed the general form of the polar perturbations as:
\begin{equation}\label{6e}
h_{\mu\nu} =
\begin{pmatrix}
 (\chi+\psi -2\eta)Y  &\zeta Y  &0  &0 \\
 \zeta Y  &(\chi+\psi)Y  &0  &0 \\
 0  &0  &\psi Y  &0 \\
 0  &0  &0  &\psi Y
\end{pmatrix},
\end{equation}
where
$ \chi $, $ \psi $, $ \zeta $ and $ \eta $ are functions of $ t $ and $ r $, and are equivalent to the gauge-invariant variables in \cite{GS} and \cite{GMG,MGG}. The polar perturbations for the Kantowski-Sachs background \eqref{1} are hence given by:
\begin{equation}\label{6f}
\begin{split}
ds^2= [1+ e \left\lbrace \chi(t,r)+\psi(t,r) -2\eta(t,r) \right\rbrace Y] dt^2 + 2e\zeta(t,r)Y dtdr 
+[-a^2(t) + e \left\lbrace \chi(t,r)+\psi(t,r) \right\rbrace Y] dr^2 \\
+[-b^2(t) +e\psi(t,r)Y] d\theta ^2
+[-b^2(t) +e \psi(t,r)Y] \sin ^2 \theta d\phi ^2  + \mathcal{O}(e^2).
\end{split}
\end{equation}
The perturbed field equations corresponding to this line element \eqref{6f} are presented in Appendix A. We consider $ \eta =0 $ for the KS metric \eqref{1} in this paper and proceed to solve the perturbation equations. In that case, the polar perturbation matrix is defined as \cite{M2}:
\begin{equation}\label{6c}
h_{\mu\nu} =
\begin{pmatrix}
 (\chi+\psi)Y  &\zeta Y  &0  &0 \\
 \zeta Y  &(\chi+\psi)Y  &0  &0 \\
 0  &0  &\psi Y  &0 \\
 0  &0  &0  &\psi Y
\end{pmatrix},
\end{equation}
and the perturbed KS line element is given by:
\begin{equation}\label{6d}
\begin{split}
ds^2= [1+ e \left\lbrace \chi(t,r)+\psi(t,r) \right\rbrace Y] dt^2 + 2e\zeta(t,r)Y dtdr 
+[-a^2(t) + e \left\lbrace \chi(t,r)+\psi(t,r) \right\rbrace Y] dr^2 \\
+[-b^2(t) +e\psi(t,r)Y] d\theta ^2
+[-b^2(t) +e \psi(t,r)Y] \sin ^2 \theta d\phi ^2  + \mathcal{O}(e^2).
\end{split}
\end{equation}
The additional term $ \eta (t,r) $ is retained in the $ (0-0) $ element of the corresponding perturbation matrix for the LTB background \cite{CL1}. For large angle fluctuations, with $ l=0 \: \textrm{or} \: 1$, the case $ \eta =0 $ does not hold for the field equations in \cite{CL1}. In the FLRW case, however, $ \eta $ must vanish \cite{M2, ROST2}. It was shown \cite{GMG,MGG} that for $l\geqslant 2$, $\eta =0$. For the polar $l=0,1$ case, gauge-invariance is not valid \cite{GMG}. Additional constraints are required for gauge fixing. For $l=1$, $\eta$ is no longer zero, so that there is one degree of gauge freedom, and the gauge-invariant variables defined for $l\geqslant 2$ become partially gauge-invariant. For $l=0$, there are two degrees of freedom (see \cite{MGG}, Appendix A).

\section{Perturbed energy-momentum tensor}

Let us consider the perturbed metric in presence of matter. The perturbations in the energy density and pressure of the fluid can be written as in the FRLW case \cite{M2, SH1, SH2} as follows:
\begin{eqnarray}
  \rho &=& \rho_0 (1+e\cdot \Delta(t,r) Y) +\mathcal{O}(e^2), \label{7a} \\
  p  &=& p_0 (1+e\cdot \Pi(t,r) Y) +\mathcal{O}(e^2), \label{7b}
\end{eqnarray}
where $ \Delta(t,r) $ and $ \Pi(t,r) $ represent the perturbations in the energy density and pressure respectively. These two perturbation terms are related because the background energy density $ \rho_0 $ and pressure $ p_0 $ are related by an equation of state.  Moreover, the fluid may or may not be co-moving with the unperturbed cosmological expansion of the universe. So perturbations in its four-velocity have to be taken into account \cite{M2}. The perturbed components of the fluid four-velocity $ u_\alpha =(u_0,u_1,u_2,u_3) $ are defined in the following way:
\begin{eqnarray}
  u_0 &=& \frac{2 g_{00}^{(0)} +e h_{00}}{2} +\mathcal{O}(e^2), \label{7c} \\
  u_1 &=& e a(t) w(t,r) Y +\mathcal{O}(e^2), \label{7d} \\
  u_2 &=& e v(t,r) (\partial_{\theta}Y) +\mathcal{O}(e^2), \label{7e} \\
  u_3 &=& = e u(t,r) \sin \theta (\partial_{\theta}Y) +\mathcal{O}(e^2). \label{7f}
\end{eqnarray}

Here, $ h_{00} $ denotes the $ (0-0) $ element of the perturbation matrix. For axial waves, $ h_{00}=0 $, whereas, in case of polar waves (with $ \eta =0 $), $ h_{00}= e(\chi+\psi)Y $. These four-velocity components satisfy the relation: $ u_{\mu}u^{\mu} = 1+ \mathcal{O}(e^2) $. Hence, the non-zero components of the perturbed energy-momentum tensor are obtained in the following form,

\underline{(i) For axial perturbations:}

\begin{eqnarray}
  T_{t t} &=& \rho_0 (1+e\Delta Y), \label{8a} \\
   T_{r r} &=& a^2 p_0(1+e  \Pi  Y) , \label{8b} \\
  T_{\theta \theta} &=& b^2 p_0(1+e  \Pi  Y)  , \label{8c} \\
  T_{\phi \phi} &=& b^2 \sin ^2 \theta p_0(1+e  \Pi  Y) , \label{8d} \\
  T_{t r} &=& (\rho_0 +p_0) e a w Y, \label{8e} \\
  T_{t \theta} &=& (\rho_0 +p_0) e v (\partial_{\theta}Y) , \label{8f} \\
  T_{t \phi} &=&  \left[ (\rho_0 +p_0)u -p_0 h_0 \right] e \sin \theta (\partial_{\theta}Y) , \label{8g} \\
  T_{r \phi} &=& -p_0 h_1 e \sin \theta (\partial_{\theta}Y). \label{8h}
\end{eqnarray}

\underline{(ii) For polar perturbations:}

\begin{eqnarray}
  T_{t t} &=& \rho_0 [1+ e(\Delta +\chi+\psi) Y], \label{9a} \\
  T_{r r} &=& p_0 [a^2 + e(a^2 \Pi -\chi-\psi)Y], \label{9b} \\
  T_{\theta \theta} &=& p_0 [b^2 + e(b^2 \Pi -\psi)Y], \label{9c} \\
  T_{\phi \phi} &=& p_0 [b^2 + e(b^2 \Pi -\psi)Y] \sin ^2 \theta, \label{9d} \\
  T_{t r} &=& e[(\rho_0 +p_0)aw -p_0\zeta]Y, \label{9e} \\
  T_{t \theta} &=& e(\rho_0 +p_0) v(\partial_{\theta}Y), \label{9f} \\
  T_{t \phi} &=& e(\rho_0 +p_0) u\sin \theta (\partial_{\theta}Y). \label{9g}
\end{eqnarray}

\section{The perturbation equations}

In this section, the perturbation equations for both axial and polar cases are presented. These are basically the linearised Einstein equations for the Kantowski-Sachs metric \eqref{1} with axial \eqref{6b} and polar \eqref{6d} perturbations. The linearised field equations for the perturbed metric in presence of matter are given by $ \mathcal{G}_{\alpha\beta} = T_{\alpha\beta} $, neglecting all terms containing second and higher orders of $ e $.

\subsection{AXIAL PERTURBATION EQUATIONS}
The linearised Einstein equations for the axially perturbed Kantowski-Sachs line element \eqref{6b} in presence of matter are represented by the set of the background field equations \eqref{2a}, \eqref{2b}, \eqref{2c} and the following set of axial perturbation equations:

\begin{equation}\label{10a}
(t-t) \hspace{0.3cm}\text{equation}:  \hspace{0.3cm}
\frac{2 \dot{a}\dot{b}}{ab} + \frac{\dot{b}^2}{b^2} +\frac{1}{b^2} =\rho_0 (1+e\Delta Y),
\end{equation}

\begin{equation}\label{10b}
(r-r) \hspace{0.3cm}\text{equation}:  \hspace{0.3cm}
-a^2\left( \frac{2\ddot{b}}{b} +\frac{{\dot{b}}^2}{b^2} +\frac{1}{b^2} \right) = a^2 p_0(1+e  \Pi  Y) ,
\end{equation}

\begin{equation}\label{10c}
(\theta-\theta) \hspace{0.3cm}\text{equation}:  \hspace{0.3cm}
-b^2\left( \frac{\ddot{a}}{a} +\frac{\ddot{b}}{b} +\frac{\dot{a} \dot{b}}{ab} \right) = b^2 p_0(1+e  \Pi  Y) ,
\end{equation}

\begin{equation}\label{10e}
(t-r) \hspace{0.3cm}\text{equation}:  \hspace{0.3cm}
(\rho_0 +p_0) e a w Y =0,
\end{equation}

\begin{equation}\label{10f}
(t-\theta) \hspace{0.3cm}\text{equation}:  \hspace{0.3cm}
(\rho_0 +p_0) e v (\partial_{\theta}Y) =0,
\end{equation}

\begin{eqnarray}\label{10g}
 (t-\phi) \hspace{0.1cm}\text{equation}:  \hspace{0.3cm} & & \frac{e}{2} \sin\theta (\partial_{\theta}Y) \left[ \frac{h_0''}{a^2} - \frac{\dot{h}_1'}{a^2}  +\frac{2 \dot{b}h_1'}{a^2 b} +\frac{2 \ddot{a}h_0}{a} +\frac{2 \ddot{b}h_0}{b} + \frac{2 \dot{a} \dot{b} h_0}{ab}
 -  \frac{h_0}{b^2 } \left\lbrace  l(l+1)  -2   \right\rbrace \right] \ \nonumber \\
   &=& \left[ (\rho_0 +p_0)u -p_0 h_0 \right] e \sin \theta (\partial_{\theta}Y),
\end{eqnarray}

\begin{eqnarray}\label{10h}
  (r-\phi) \hspace{0.1cm}\text{equation}:  \hspace{0.3cm} & & -\frac{e}{2} \sin\theta (\partial_{\theta}Y) \left[
\ddot{h}_1 -\dot{h}_0' -\frac{\dot{a}\dot{h}_1}{a}  +\frac{\dot{a}h_0'}{a} -\frac{2 \dot{b} h_0'}{b} -\frac{2\ddot{a} h_1}{a} -\frac{4\ddot{b} h_1}{b} -\frac{2 \dot{b}^2 h_1}{b^2}
+ \frac{h_1}{b^2} \left\lbrace l(l+1)  -2   \right\rbrace  \right] \nonumber \\
   &=& -p_0 h_1 e \sin \theta (\partial_{\theta}Y),
\end{eqnarray}

\begin{equation}\label{10i}
(\theta-\phi) \hspace{0.3cm}\text{equation}:  \hspace{0.3cm}
 \frac{e}{2} \left( \dot{h}_0+ \frac{\dot{a}h_0}{a} -\frac{h_1'}{a^2} \right) \left( \cos\theta (\partial_\theta Y)-\frac{2\sin \theta(\partial_\theta Y)}{\theta} +\sin\theta (\partial_\theta \partial_\theta Y) \right) =0.
\end{equation}

Upon simplifying and using the background field equations \eqref{2a}-\eqref{2c}, these equations lead to the following set of equations :
\begin{eqnarray}
  \Delta \cdot \rho_0 &=& 0, \label{11a} \\
  \Pi  \cdot p_0 &=& 0, \label{11b} \\
  w (\rho_0 +p_0) &=& 0, \label{11c} \\
  v (\rho_0 +p_0) &=& 0, \label{11d} \\
  \frac{h_0''}{a^2} -\frac{\dot{h}_1'}{a^2}  +\frac{2 \dot{b}h_1'}{a^2 b}
-\frac{h_0}{b^2} \left\lbrace l(l+1)  -2  \right\rbrace &=& 2u(\rho_0 +p_0) , \label{11e} \\
  \ddot{h}_1 -\dot{h}_0' -\frac{\dot{a}\dot{h}_1}{a}  +\frac{\dot{a}h_0'}{a} -\frac{2 \dot{b} h_0'}{b} -\frac{2\ddot{a} h_1}{a}
 +\frac{h_1}{b^2} \left\lbrace l(l+1)  -2  \right\rbrace  &=& 0, \label{11f} \\
  \dot{h}_0+ \frac{\dot{a}h_0}{a} -\frac{h_1'}{a^2}  &=& 0. \label{11g}
\end{eqnarray}

The equations \eqref{11a}-\eqref{11d} imply that $ \Delta =  \Pi =w =v =0 $. Thus the matter perturbations disappear even in the presence of matter. The only perturbing term that remains is the azimuthal velocity $ u $ in equation \eqref{11e}. This ensures that the axial waves do not perturb the matter distribution but can trigger perturbations in only the azimuthal component of the fluid velocity. From equations \eqref{11c} and \eqref{11d}, we can conclude that either both $ w $ and $ v $ vanish or $ p_0 = -\rho_0 $. The latter choice depends on the equation of state and may be investigated further, in which case $ w $ and $ v $ will possibly be non-zero.

To make the equations simpler and their analytical solutions feasible, we choose to work with the perturbed spacetime in vacuum. In absence of matter, the axial perturbation equations reduce to:
\begin{equation} \label{12a}
\frac{h_0''}{a^2} -\frac{\dot{h}_1'}{a^2}  +\frac{2 \dot{b}h_1'}{a^2 b} -\frac{h_0}{b^2 } \left\lbrace  l(l+1)-2 \right\rbrace  =0,
\end{equation}
\begin{equation} \label{12b}
\ddot{h}_1 -\dot{h}_0' -\frac{\dot{a}\dot{h}_1}{a}  +\frac{\dot{a}h_0'}{a} -\frac{2 \dot{b} h_0'}{b} -\frac{2\ddot{a} h_1}{a} -\frac{4\ddot{b} h_1}{b} -\frac{2 \dot{b}^2 h_1}{b^2} +\frac{h_1}{b^2} \left\lbrace  l(l+1)-2 \right\rbrace =0,
\end{equation}
and
\begin{equation}\label{12c}
\dot{h}_0 +\frac{\dot{a}h_0}{a} -\frac{h_1'}{a^2} =0
\hspace{0.5cm} \Rightarrow {h_1'}= a^2\dot{h}_0+ a\dot{a}h_0.
\end{equation}
We are going to solve these equations in Sec. VIA.

\subsection{POLAR PERTURBATION EQUATIONS}
The following linearised field equations hold for the polar-perturbed metric \eqref{6d} in presence of matter :

\begin{equation}\label{13a}
\begin{split}
(t-t) \hspace{0.3cm} \text{equation}:  \hspace{0.3cm}
\dfrac{4 \sin \theta}{4[-b^2+ e\psi Y]^3 [-a^2+ eY(1- a^2)(\chi+\psi)]^2 \sin \theta} \cdot \left[(a^4b^4 + a^4b^4\dot{b}^2 + 2a^3\dot{a}b^5\dot{b})
\right. \\ \left.
+e \left\lbrace \frac{1}{2}a^2b^2(b^2\chi +b^2\psi +a^2\psi) \left( \cot \theta (\partial_{\theta}Y) + (\partial_{\theta}\partial_{\theta}Y) \right) \right\rbrace
+eY \left\lbrace (3a^4b^4 - 2a^2b^4 + 4a^3\dot{a}b^5\dot{b}
\right. \right. \\ \left. \left.
- 2a\dot{a}b^5\dot{b} - 2a^2b^4\dot{b}^2 + 2a^4b^4\dot{b}^2)(\chi+\psi) - (2a^4b^2 + 4a^3\dot{a}b^3\dot{b} + a^4b^2\dot{b}^2)\psi - a^2b^5\dot{b}(\dot{\chi}+\dot{\psi})
\right. \right. \\ \left. \left.
 - (a^4b^3\dot{b} + a^3\dot{a}b^4)\dot{\psi} +a^2b^4\psi'' +2a^2b^5\dot{b}\zeta'  \right\rbrace  \right]
=  \rho_0 [1+ e(\Delta +\chi+\psi) Y]
\end{split}
\end{equation}

\begin{equation}\label{13b}
\begin{split}
(r-r) \hspace{0.3cm}\text{equation}:  \hspace{0.3cm}
\dfrac{4 \sin \theta}{4[b^2- e\psi Y]^3 [-a^2+ eY(1- a^2)(\chi+\psi)]^2 \sin \theta} \cdot \left[(-2a^6b^5\ddot{b} - a^6b^4\dot{b}^2 - a^6b^4)
\right. \\ \left.
+e \left\lbrace \frac{1}{2}a^6b^2(b^2\chi +b^2\psi -\psi) \left( \cot \theta (\partial_{\theta}Y) + (\partial_{\theta}\partial_{\theta}Y) \right) \right\rbrace
+eY \left\lbrace (- a^6b^4\dot{b}^2 - 2a^6b^4 +3a^4b^4 + 6a^4b^5\ddot{b}
\right. \right. \\ \left. \left.
+ 3a^4b^4\dot{b}^2 - 2a^6b^5\ddot{b} )(\chi+\psi) + (4a^6b^3\ddot{b} + 3a^6b^2\dot{b}^2 + 2a^6b^2)\psi + a^6b^5\dot{b}(\dot{\chi}+\dot{\psi})
 - a^6b^3\dot{b}\dot{\psi} + a^6b^4\ddot{\psi}  \right\rbrace  \right]  \\
= p_0 [a^2 + e(a^2 \Pi -\chi-\psi)Y],
\end{split}
\end{equation}

\begin{equation}\label{13c}
\begin{split}
(\theta-\theta) \hspace{0.3cm}\text{equation}:  \hspace{0.3cm}
\dfrac{-4 \sin \theta}{4[-b^2 +e\psi Y] [-a^2+ eY(1- a^2)(\chi+\psi)]^2 \sin \theta} \cdot \left[(-a^3\ddot{a}b^4 - a^4b^3\ddot{b} - a^3\dot{a}b^3\dot{b})
\right. \\ \left.
+e \left\lbrace \frac{1}{2}a^2b^2(a^2 -1)(\chi+\psi) \left( \cot \theta (\partial_{\theta}Y)  \right)
+ (a\ddot{a}b^4 - a^4b^3\ddot{b} - a^3\ddot{a}b^4 - a^3\dot{a}b^3\dot{b} + a\dot{a}b^3\dot{b} + 2a^2b^3\ddot{b}
\right. \right. \\ \left. \left.
 +\dot{a}^2b^4)(\chi+\psi) + (a^4\dot{b}^2 + a^4b\ddot{b} + a^3\dot{a}b\dot{b} + 2a^3\ddot{a}b^2)\psi
 + (-a\dot{a}b^4 + \frac{1}{2}a^3\dot{a}b^4 + \frac{1}{2}a^4b^3\dot{b} + \frac{1}{2}a^2b^3\dot{b})(\dot{\chi}+\dot{\psi})
\right. \right. \\ \left. \left.
 + ( - a^4b\dot{b} + \frac{1}{2}a^3\dot{a}b^2)\dot{\psi} + \frac{1}{2} a^2b^4(\ddot{\chi}+\ddot{\psi}+\chi''+\psi'') + \frac{1}{2}a^4b^2 \ddot{\psi} - \frac{1}{2}a^2b^2 \psi'' - a^2b^3\dot{b}\zeta' - a^2b^4\dot{\zeta}' \right\rbrace \right]  \\
=  p_0 [b^2 + e(b^2 \Pi -\psi)Y],
\end{split}
\end{equation}

\begin{equation}\label{13d}
\begin{split}
(\phi-\phi) \hspace{0.3cm}\text{equation}:  \hspace{0.3cm}
 \dfrac{-4 \sin ^2 \theta}{4[-b^2 +e\psi Y] [-a^2+ eY(1- a^2)(\chi+\psi)]^2} \cdot \left[(-a^3\ddot{a}b^4 - a^4b^3\ddot{b} - a^3\dot{a}b^3\dot{b})
\right. \\ \left.
+e \left\lbrace \frac{1}{2}a^2b^2(a^2 -1)(\chi+\psi) (\partial_{\theta}\partial_{\theta}Y)
+ (a\ddot{a}b^4 - a^4b^3\ddot{b} - a^3\ddot{a}b^4 - a^3\dot{a}b^3\dot{b} + a\dot{a}b^3\dot{b} + 2a^2b^3\ddot{b}
\right. \right. \\ \left. \left.
 +\dot{a}^2b^4)(\chi+\psi) + (a^4\dot{b}^2 + a^4b\ddot{b} + a^3\dot{a}b\dot{b} + 2a^3\ddot{a}b^2)\psi
 + (-a\dot{a}b^4 + \frac{1}{2}a^3\dot{a}b^4 + \frac{1}{2}a^4b^3\dot{b} + \frac{1}{2}a^2b^3\dot{b})(\dot{\chi}+\dot{\psi})
 \right. \right. \\ \left. \left.
 + ( - a^4b\dot{b} + \frac{1}{2}a^3\dot{a}b^2)\dot{\psi}
+ \frac{1}{2} a^2b^4(\ddot{\chi}+\ddot{\psi}+\chi''+\psi'') + \frac{1}{2}a^4b^2 \ddot{\psi} - \frac{1}{2}a^2b^2 \psi'' - a^2b^3\dot{b}\zeta' - a^2b^4\dot{\zeta}'  \right\rbrace \right]\\
=  p_0 [b^2 + e(b^2 \Pi -\psi)Y] \sin ^2 \theta,
\end{split}
\end{equation}

\begin{equation}\label{13e}
\begin{split}
(t-r) \hspace{0.3cm} \text{equation}:  \hspace{0.3cm}
\dfrac{ea^4b^4 \sin \theta}{[-b^2+ e\psi Y]^3 [a^2- eY(1- a^2)(\chi+\psi)]^2 \sin \theta} \cdot \left[ \left\lbrace  -\dot{\psi}' +\left(\frac{\dot{a}}{a} + \frac{\dot{b}}{b}\right) \psi' - b\dot{b}(\chi'+\psi')
\right. \right. \\ \left. \left.
-(2b\ddot{b} + \dot{b}^2) \zeta \right\rbrace Y
-\frac{\zeta}{2} \left\lbrace (\partial_{\theta} \partial_{\theta} Y) + \cot \theta (\partial_{\theta} Y)\right\rbrace  \right]
=  e[(\rho_0 +p_0)aw -p_0\zeta]Y,
\end{split}
\end{equation}

\begin{equation}\label{13f}
\begin{split}
(t-\theta) \hspace{0.3cm} \text{equation}:  \hspace{0.3cm}
\dfrac{e (\partial_{\theta} Y)}{2[b^2- e\psi Y]^2 [a^2- eY(1- a^2)(\chi+\psi)]^2} \cdot \left[(-a\dot{a}b^4 + a^3\dot{a}b^4 - a^2b^3\dot{b} + a^4b^3\dot{b})(\chi+\psi)
 \right. \\ \left.
-2a^4b\dot{b}\psi +a^2b^4(\dot{\chi}+\dot{\psi}-\zeta') +a^4b^2\dot{\psi} \right]
= e(\rho_0 +p_0) v(\partial_{\theta}Y),
\end{split}
\end{equation}

\begin{equation}\label{13g}
(r-\theta) \hspace{0.3cm} \text{equation}:  \hspace{0.3cm}
 \dfrac{e (\partial_{\theta} Y) a^4b^4}{2[-b^2+ e\psi Y]^2 [a^2- eY(1- a^2)(\chi+\psi)]^2} \cdot \left[-(\chi' +\psi' -\dot{\zeta})
\right. \\ \left.
+ \frac{1}{b^2}\psi' +\frac{\dot{a}}{a}\zeta  \right] =0,
\end{equation}

\begin{equation}\label{13h}
(t-\phi) \hspace{0.3cm} \text{equation}:  \hspace{0.3cm}
e(\rho_0 +p_0) u\sin \theta (\partial_{\theta}Y) =0,
\end{equation}

It is to be noted here that the above equations correspond to $ \eta =0 $. These equations are simplified by substituting the background field equations \eqref{2a}-\eqref{2c}, using the relation \eqref{sph}, and dropping all terms containing second order or higher powers of $ e $ in the expansions. We are therefore left with the following set of equations :

\begin{equation}\label{14a}
w= \frac{1}{ab^2 (\rho_0+p_0)} \left[ - \dfrac{l(l+1)}{2} \zeta + b\dot{b} \chi' +\left(-\frac{\dot{a}}{a} - \frac{\dot{b}}{b} + b\dot{b}\right) \psi' + \dot{\psi}' \right],
\end{equation}

\begin{equation}\label{14b}
v= \frac{1}{2(\rho_0+p_0)} \left[ \left( -\frac{\dot{a}}{a^3} + \frac{\dot{a}}{a} - \frac{\dot{b}}{a^2b} + \frac{\dot{b}}{b} \right)\chi +\frac{1}{a^2}\dot{\chi}
+\left( -\frac{\dot{a}}{a^3} + \frac{\dot{a}}{a} - \frac{\dot{b}}{a^2b} + \frac{\dot{b}}{b} - \frac{2\dot{b}}{b^3}\right)\psi
+ \left( \frac{1}{a^2}+\frac{1}{b^2}\right) \dot{\psi} - \frac{1}{a^2}\zeta' \right],
\end{equation}

\begin{equation}\label{14c}
u =0,
\end{equation}

\begin{equation}\label{14d}
\frac{\dot{a}}{a}\zeta +\dot{\zeta} -\chi' +\left(\frac{1}{b^2}-1\right) \psi' =0,
\end{equation}

\begin{equation}\label{14e}
\begin{split}
\Delta = \frac{1}{a^4b^6\rho_0} \left[ \left\lbrace -a^4b^4\dot{b}^2 - 2a^3\dot{a}b^5\dot{b} + 2a\dot{a}b^5\dot{b}
-\frac{l(l+1)}{2} \cdot a^2b^4 \right\rbrace \chi
 + \left\lbrace a^4b^2 - a^4b^4\dot{b}^2 + 2a^4b^2\dot{b}^2 -2a^3\dot{a}b^5\dot{b}
\right. \right. \\ \left. \left.
+ 2a\dot{a}b^5\dot{b} + 2a^3\dot{a}b^3\dot{b} -\frac{l(l+1)}{2} \cdot a^2b^2(a^2+b^2) \right\rbrace \psi
-a^2b^5\dot{b}\dot{\chi} - \left\lbrace a^2b^5\dot{b} + a^4b^3\dot{b} + a^3\dot{a}b^4 \right\rbrace \dot{\psi}
+ a^2b^4\psi'' +2a^2b^5\dot{b}\zeta'  \right],
\end{split}
\end{equation}

\begin{equation}\label{14f}
\begin{split}
\Pi = \frac{1}{a^6b^6p_0} \left[ \left\lbrace a^6b^4\dot{b}^2 + 2a^6b^5\ddot{b} - \frac{l(l+1)}{2} \cdot a^6b^4 \right\rbrace \chi  \qquad\qquad\qquad\qquad \right. \\ \left.  + \left\lbrace -a^6b^2 + a^6b^4\dot{b}^2 - 2a^6b^3\ddot{b} + 2a^6b^5\ddot{b} - \frac{l(l+1)}{2} \cdot a^6b^2(b^2 -1)\right\rbrace \psi
\right. \\ \left. + a^6b^5\dot{b}\dot{\chi} + (a^6b^5\dot{b} - a^6b^3\dot{b})\dot{\psi} + a^6b^4\ddot{\psi} \right],
\end{split}
\end{equation}

\begin{equation}\label{14g}
\begin{split}
p_0 b^2\Pi  = \frac{1}{a^4b^2} \left[ (-a\ddot{a}b^4 + a^4b^3\ddot{b} + a^3\ddot{a}b^4 + a^3\dot{a}b^3\dot{b}
- a\dot{a}b^3\dot{b} + \dot{a}^2b^4)(\chi+\psi) + (a^4\dot{b}^2 - a^4 b\ddot{b} -  a^3\dot{a}b\dot{b})\psi
\right. \\ \left.
+ (-a\dot{a}b^4 + \frac{1}{2}a^3\dot{a}b^4 + \frac{1}{2}a^4b^3\dot{b}
+ \frac{1}{2}a^2b^3\dot{b})(\dot{\chi}+\dot{\psi})
+ ( - a^4b\dot{b} + \frac{1}{2}a^3\dot{a}b^2)\dot{\psi} + \frac{1}{2} a^2b^4(\ddot{\chi}+\ddot{\psi}+\chi''+\psi'')
\right. \\ \left.
+ \frac{1}{2}a^4b^2 \ddot{\psi} - \frac{1}{2}a^2b^2 \psi'' - a^2b^3\dot{b}\zeta' - a^2b^4\dot{\zeta}' + \frac{1}{2Y}a^2b^2(a^2 -1)(\chi+\psi) \cot \theta (\partial_{\theta}Y)
\right],
\end{split}
\end{equation}

\begin{equation}\label{14h}
\begin{split}
p_0 b^2\Pi  = \frac{1}{a^4b^2} \left[ (-a\ddot{a}b^4 + a^4b^3\ddot{b} + a^3\ddot{a}b^4 + a^3\dot{a}b^3\dot{b}
- a\dot{a}b^3\dot{b} + \dot{a}^2b^4)(\chi+\psi) + (a^4\dot{b}^2 - a^4 b\ddot{b} -  a^3\dot{a}b\dot{b})\psi
\right. \\ \left.
 + (-a\dot{a}b^4 + \frac{1}{2}a^3\dot{a}b^4 + \frac{1}{2}a^4b^3\dot{b}
+ \frac{1}{2}a^2b^3\dot{b})(\dot{\chi}+\dot{\psi})
+ ( - a^4b\dot{b} + \frac{1}{2}a^3\dot{a}b^2)\dot{\psi} + \frac{1}{2} a^2b^4(\ddot{\chi}+\ddot{\psi}+\chi''+\psi'')
\right. \\ \left.
+ \frac{1}{2}a^4b^2 \ddot{\psi} - \frac{1}{2}a^2b^2 \psi'' - a^2b^3\dot{b}\zeta' - a^2b^4\dot{\zeta}' + \frac{1}{2Y}a^2b^2(a^2 -1)(\chi+\psi) (\partial_{\theta} \partial_{\theta}Y) \right],
\end{split}
\end{equation}

Adding equation \eqref{14g} to \eqref{14h} and substituting the expression for $\Pi$ from equation \eqref{14f} in the resulting equation, we get
\begin{equation}\label{14i}
\begin{split}
\left[ -a\ddot{a}b^4 - a^4b^3\ddot{b} + a^3\ddot{a}b^4 + a^3\dot{a}b^3\dot{b} - a\dot{a}b^3\dot{b} + \dot{a}^2b^4 -a^4b^2\dot{b}^2
+ \frac{l(l+1)}{4} \cdot (a^4b^2 +a^2b^2) \right] \chi \\
+ \left[ -a\ddot{a}b^4 - a^4b^3\ddot{b} + a^3\ddot{a}b^4 + a^3\dot{a}b^3\dot{b} - a\dot{a}b^3\dot{b} - a^3\dot{a}b\dot{b} + \dot{a}^2b^4 + a^4\dot{b}^2 + a^4 b\ddot{b} - a^4b^2\dot{b}^2 +a^4
\right. \\ \left.
+ \frac{l(l+1)}{4} \cdot (a^4b^2 +a^2b^2 -2a^4) \right]  \psi + \left[ -a\dot{a}b^4 + \frac{1}{2}a^3\dot{a}b^4 - \frac{1}{2}a^4b^3\dot{b}
+ \frac{1}{2}a^2b^3\dot{b} \right] \dot{\chi}\\
+ \left[ -a\dot{a}b^4 + \frac{1}{2}a^3\dot{a}b^4 - \frac{1}{2}a^4b^3\dot{b} + \frac{1}{2}a^2b^3\dot{b} + \frac{1}{2}a^3\dot{a}b^2 - a^4b^3\dot{b} \right] \dot{\psi}
+ \frac{1}{2} a^2b^4 \left(\ddot{\chi}+\chi''\right) \\
+ \left(\frac{1}{2}a^2b^4 - \frac{1}{2}a^4b^2\right) \ddot{\psi} + \left(\frac{1}{2}a^2b^4 - \frac{1}{2}a^2b^2\right) \psi'' - a^2b^3\dot{b}\zeta' - a^2b^4\dot{\zeta}' =0 .
\end{split}
\end{equation}
Equations \eqref{14a} and \eqref{14b} represent the deformation of the two components of the fluid four-velocity. It follows from equation \eqref{14c} that the remaining component, i.e. the azimuthal velocity $ u $ is zero in case of polar waves. This feature of the polar waves contradicts that of the axial waves as already observed in the preceding subsection. Besides, the polar waves bring about perturbations in the energy density and pressure, which are determined by equations \eqref{14e} and \eqref{14f} respectively. Equations \eqref{14d} and \eqref{14i} do not explicitly depend on matter. The three master equations (following Clarkson \textit{et al.} \cite{CL1}) governing the evolution of the three perturbing variables in Kantowski-Sachs background are the equations \eqref{14d}, \eqref{14f} and \eqref{14i}, together with the constraint equation $ \eta =0 $, while equations \eqref{14a}, \eqref{14b} and \eqref{14e} act as constraints. If the matter perturbations vanish, i.e. $ \Delta = \Pi = w = v =0 $, then from equations \eqref{14a}, \eqref{14b}, \eqref{14e} and \eqref{14f}, it can be inferred that $ \chi = \psi = \zeta =0 $, which implies that the polar modes are absent. In other words, travelling polar GWs must leave inhomogeneous and anisotropic imprints in the background spacetime.

In absence of matter, the set of polar perturbation equations \eqref{13a}-\eqref{13h} become somewhat less involved. These equations (with $ \eta =0 $) are presented below:
\begin{equation}\label{15a}
- \dfrac{l(l+1)}{2} \zeta + b\dot{b} \chi' +\left(-\frac{\dot{a}}{a} - \frac{\dot{b}}{b} + b\dot{b}\right) \psi' + \dot{\psi}' =0,
\end{equation}

\begin{equation}\label{15b}
\begin{split}
\left( -\frac{\dot{a}}{a^3} + \frac{\dot{a}}{a} - \frac{\dot{b}}{a^2b} + \frac{\dot{b}}{b} \right)\chi +\frac{1}{a^2}\dot{\chi}
+\left( -\frac{\dot{a}}{a^3} + \frac{\dot{a}}{a} - \frac{\dot{b}}{a^2b} + \frac{\dot{b}}{b} - \frac{2\dot{b}}{b^3}\right)\psi
+ \left( \frac{1}{a^2}+\frac{1}{b^2}\right) \dot{\psi} - \frac{1}{a^2}\zeta'  =0,
\end{split}
\end{equation}

\begin{equation}\label{15c}
\frac{\dot{a}}{a}\zeta +\dot{\zeta}-\chi' +\left(\frac{1}{b^2}-1\right) \psi' =0,
\end{equation}

\begin{equation}\label{15d}
\left\lbrace a^2 + \frac{2\dot{a}b\dot{b}}{a} -\frac{l(l+1)}{2} \right\rbrace \chi
+ \left\lbrace  a^2 + \frac{2\dot{a}b\dot{b}}{a}  + \frac{a^2 \dot{b}^2}{b^2}
 -\frac{l(l+1)}{2} \cdot\left( 1+\frac{  a^2  }{b^2}\right) \right\rbrace \psi - b\dot{b}\dot{\chi} - \left\lbrace \frac{a^2\dot{b}}{b} +  a\dot{a}  + b\dot{b} \right\rbrace \dot{\psi} + \psi'' +2b\dot{b}\zeta'  =0,
\end{equation}

\begin{equation}\label{15e}
\left\lbrace 1  +  \frac{l(l+1)}{2} \right\rbrace \chi  + \left\lbrace 1 - \frac{\dot{b}^2}{b^2}  +  \frac{l(l+1)}{2} \cdot\left( 1- \frac{1}{b^2}\right) \right\rbrace \psi
- b\dot{b}\dot{\chi} + \left( \frac{\dot{b}}{b} -b\dot{b} \right) \dot{\psi} - \ddot{\psi} =0,
\end{equation}

\begin{equation}\label{15f}
\begin{split}
(\dot{a}^2b^4 + a^2b^3\ddot{b})(\chi+\psi) + (a^4\dot{b}^2 + a^3\ddot{a}b^2)\psi
+ (-a\dot{a}b^4 + \frac{1}{2}a^3\dot{a}b^4 + \frac{1}{2}a^4b^3\dot{b} + \frac{1}{2}a^2b^3\dot{b})(\dot{\chi}+\dot{\psi}) \\
+ ( - a^4b\dot{b} + \frac{1}{2}a^3\dot{a}b^2)\dot{\psi} + \frac{1}{2} a^2b^4(\ddot{\chi}+\ddot{\psi}+\chi''+\psi'')
+ \frac{1}{2}a^4b^2 \ddot{\psi} - \frac{1}{2}a^2b^2 \psi''  \\
 - a^2b^3\dot{b}\zeta' - a^2b^4\dot{\zeta}' + \frac{1}{2Y}a^2b^2(a^2 -1)(\chi+\psi) \cot \theta (\partial_{\theta}Y) =0,
\end{split}
\end{equation}

\begin{equation}\label{15g}
\begin{split}
(\dot{a}^2b^4 + a^2b^3\ddot{b})(\chi+\psi) + (a^4\dot{b}^2 + a^3\ddot{a}b^2)\psi
+ (-a\dot{a}b^4 + \frac{1}{2}a^3\dot{a}b^4 + \frac{1}{2}a^4b^3\dot{b} + \frac{1}{2}a^2b^3\dot{b})(\dot{\chi}+\dot{\psi}) \\
+ ( - a^4b\dot{b} + \frac{1}{2}a^3\dot{a}b^2)\dot{\psi} + \frac{1}{2} a^2b^4(\ddot{\chi}+\ddot{\psi}+\chi''+\psi'')
+ \frac{1}{2}a^4b^2 \ddot{\psi} - \frac{1}{2}a^2b^2 \psi''  \\
 - a^2b^3\dot{b}\zeta' - a^2b^4\dot{\zeta}' + \frac{1}{2Y}a^2b^2(a^2 -1)(\chi+\psi) (\partial_{\theta} \partial_{\theta}Y) =0.
\end{split}
\end{equation}

Adding equations \eqref{15f} and \eqref{15g}, and simplifying gives
\begin{equation}\label{15h}
\begin{split}
\left\lbrace b\ddot{b} + \frac{\dot{a}^2 b^2}{a^2} - \frac{l(l+1)}{  4 }  (a^2 -1) \right\rbrace \chi
+ \left\lbrace a\ddot{a} + b\ddot{b} + \frac{\dot{a}^2 b^2}{a^2} + \frac{a^2 \dot{b}^2}{b^2} - \frac{l(l+1)}{  4  } (a^2 -1) \right\rbrace \psi \\
+ \left\lbrace - \frac{\dot{a} b^2}{a} +\frac{1}{2} b\dot{b} +\frac{1}{2} a\dot{a} b^2 +\frac{1}{2} a^2 b\dot{b} \right\rbrace \dot{\chi}
+ \left\lbrace - \frac{\dot{a} b^2}{a} - \frac{a^2 \dot{b}}{b} + \frac{1}{2} a\dot{a} +\frac{1}{2} b\dot{b} +\frac{1}{2} a\dot{a} b^2 +\frac{1}{2} a^2 b\dot{b} \right\rbrace  \dot\psi \\
+ \frac{1}{2} b^2(\ddot{\chi}+\chi'') + \frac{1}{2} (a^2+b^2) \ddot{\psi} + \frac{1}{2} (b^2 -1) \psi''
- b\dot{b}\zeta' - b^2\dot{\zeta}' =0
\end{split}
\end{equation}
We will discuss the solutions for these equations in Sec. VIB.

\section{Gravitational waves in Kantowski-Sachs background}
We now solve the axial and polar perturbation equations in order to ascertain the nature of the GWs propagating through the Kantowski-Sachs universe.

\subsection{AXIAL GRAVITATIONAL WAVES}
We proceed to solve the axial perturbation equations to get the expressions for $ h_0(t,r) $ and $ h_1(t,r) $. We attempt to find the solutions analytically in the vacuum case where the equations become simpler to handle. It is to be noted that \eqref{12c} is a constraint equation connecting the two axial perturbation parameters. It enables decoupling of the equation \eqref{12a} for $ h_0 (t,r) $ from $ h_1 (t,r) $ and simplifies the solution finding process. Substituting equation \eqref{12c} in equation \eqref{12a} leads to: 
\begin{equation} \label{16a}
-\ddot{h}_0 +\frac{h_0''}{a^2} -\frac{3\dot{a}\dot{h}_0}{a} +\frac{2\dot{b}\dot{h}_0}{b} -\frac{\ddot{a} h_0}{a} -\frac{\dot{a}^2 h_0}{a^2} +\frac{2 \dot{a}\dot{b}h_0}{ab} -\frac{h_0}{b^2} \left\lbrace l(l+1)-2 \right\rbrace =0.
\end{equation}
Eliminating $ a(t) $ from equation \eqref{16a} with the help of equation \eqref{5hi}, we arrive at
\begin{equation}\label{16b}
 - \ddot{h}_0 +b h_0''  +\frac{7\dot{b}}{2b} \dot{h}_0 -\frac{2{\dot{b}}^2}{b^2} h_0  + \frac{\ddot{b}}{2b}  h_0
-\frac{1}{b^2} \left\lbrace l(l+1)-2 \right\rbrace h_0 =0. 
\end{equation}

Now, let us define a new quantity $ \mathcal{Q}(t,r) $ such that
\begin{equation} \label{AG1}
h_0(t,r)= r^\alpha (b(t))^\beta \mathcal{Q}(t,r).
\end{equation}
Here $ \alpha $ and $ \beta $ can take any integral or fractional values. Inserting the expression of $ b(t) $ from equation \eqref{5hi}, and also choosing  $ l=2 $, equation \eqref{16b} reduces to
\begin{equation} \label{AG2}
\begin{split}
- K^\beta t^{\frac{2\beta}{3}} r^\alpha \ddot{\mathcal{Q}} + K^{\beta +1} t^{\frac{2}{3}(\beta +1)} r^\alpha \mathcal{Q}''  + \left[ \left( -\frac{4}{3}\beta + \frac{7}{3}\right) K^\beta t^{(\frac{2\beta}{3}-1)} r^\alpha \right]  \dot{\mathcal{Q}}
+ 2\alpha K^{\beta +1} t^{\frac{2}{3}(\beta +1)} r^{\alpha -1}  \mathcal{Q}' \\
+ \left[ \alpha^2 K^{\beta +1}  t^{\frac{2}{3}(\beta +1)} r^{\alpha -2}  -  \alpha K^{\beta +1} t^{\frac{2}{3}(\beta +1)} r^{\alpha -2}  +  \left( \frac{20}{9} \beta -  \frac{4}{9} \beta^2 -1 \right) K^\beta t^{(\frac{2\beta}{3}-2)} r^\alpha
- 4 K^{\beta-2} t^{(\frac{2\beta}{3}-\frac{4}{3})} r^\alpha \right] \mathcal{Q}  =0.
\end{split}
\end{equation}

This is a wave equation in $ \mathcal{Q}(t,r) $. The master equation \eqref{AG2} describes one of the two gravitational modes and can be solved independently of the remaining equations. It can be solved by the method of separation of variables by assuming
\begin{equation}\label{AG3}
\mathcal{Q}(t,r)= \mathcal{T}(t) \mathcal{R}(r).
\end{equation}

The equation \eqref{AG2} then yields two differential equations with the separation constant $ -m^2 $ :
\begin{equation}\label{AG4}
\ddot{\mathcal{T}} + \left( \frac{4\beta}{3} - \frac{7}{3}\right)  \frac{\dot{\mathcal{T}}}{t} +\left( -\frac{20\beta}{9t^{2}}  + \frac{4\beta^2}{9t^{2}} + \frac{1}{t^{2}} +\frac{4}{K^2 t^{4/3}} + m^2 K t^{2/3} \right)\mathcal{T}  =0,
\end{equation}
\begin{equation}\label{AG5}
\text{and} \hspace{0.5cm}  {\mathcal{R}''} +  \frac{2\alpha \mathcal{R}'}{r} +\left( \frac{\alpha^2}{r^2} - \frac{\alpha}{r^2} + m^2 \right) \mathcal{R} =0.
\end{equation}

The solutions are respectively:
\begin{equation}\label{AG6}
\mathcal{T}(t) = c_1 t^{(3- 2\beta/3)} {\exp(\xi t^{4/3})} \Sigma  +c_2 t^{(3- 2\beta/3)} \exp(\xi t^{4/3})
 \Sigma  \left( \int \dfrac{\exp(-2\xi t^{4/3}) dt }{t^{11/3} \Sigma^2} \right) ,
\end{equation}
\begin{equation}\label{AG7}
\text{and} \hspace{0.5cm}  \mathcal{R}(r) = r^{-\alpha} \left( c_3 \sin mr + c_4 \cos mr \right) .
\end{equation}
Here, $ c_1 $, $ c_2 $, $ c_3 $ and $ c_4 $ are the integration constants, \hspace{0.1cm} $ \xi =\dfrac{3}{4} \sqrt{-K m^2} $, and
$ \Sigma =\text{HeunB}\left( 4, 0, 0, \dfrac{6\sqrt{6}i}{K^2 (-Km^2)^{1/4}}, \dfrac{\sqrt{6}i}{2} (-K m^2)^{1/4} t^{2/3} \right)  $ represent the Heun function which is the solution for Heun's biconfluent equation \cite{HOR, RON}. Thus the expression for $ \mathcal{Q}(t,r) $ can be obtained in terms of $ \mathcal{T}(t) $ and $ \mathcal{R}(r) $. Substituting it in equation \eqref{AG1} and also using equation \eqref{5hi}, one can find the expression for $ h_0(t,r) $ : 
\begin{equation}\label{AG8}
h_0(t,r)= r^\alpha K^\beta  t^{2\beta/3} \mathcal{Q}(t,r).
\end{equation}

Finally, $ h_1(t,r) $ is determined from equation \eqref{12c} as:
\begin{equation}\label{AG9}
h_1(t,r)=  f(t)+ \frac{1}{K t^{2/3}} \int_{r_0}^{r}\left(   \dot{h}_0(t,r) - \frac{h_0(t,r)}{3t} \right) dr ,
\end{equation}
where $ f(t) $ is the integration constant, and $ r_0 $ characterizes the initial hypersurface giving rise to gravitational waves. The function $ f(t) $ is arbitrary, and can be chosen to be zero, $ f(t) = 0$, and it follows that
\begin{equation}\label{AG10}
h_1(t,r) = K^{\beta-1}\left(  t^{(\frac{2\beta}{3}-\frac{2}{3})} \dot{\mathcal{T}} + \left( \frac{2}{3} \beta -\frac{1}{3}\right) t^{(\frac{2\beta}{3}-\frac{5}{3})} \mathcal{T} \right) \int_{r_0}^{r}  r^{\alpha} \mathcal{R} dr.
\end{equation}

For the purpose of illustration, we present a few cases by choosing specific values of $ \alpha $ and $ \beta $.

\bigskip

\underline{\textbf{Case I: $ \alpha =1 $ and $ \beta =2 $}}

\bigskip

In this case equation \eqref{AG1} becomes:
\begin{equation} \label{17a}
h_0(t,r)= r b(t)^2 \mathcal{Q}(t,r).
\end{equation}
On substituting this expression, equation \eqref{AG2} reads:
\begin{equation} \label{17b}
-r b^2 \ddot{\mathcal{Q}} +r b^3 \mathcal{Q}'' -\frac{1}{2}r b\dot{b} \dot{\mathcal{Q}} +2b^3 \mathcal{Q}'  -\frac{3}{2}r b \ddot{b} \mathcal{Q} +3r \dot{b}^2 \mathcal{Q} -r \left\lbrace l(l+1)-2 \right\rbrace \mathcal{Q} =0.
\end{equation}
Inserting the expression of $ b(t) $ from equation \eqref{5hi}, and also choosing  $ l=2 $, the wave equation \eqref{17b} in $ \mathcal{Q}(t,r) $ reduces to:
\begin{equation} \label{17c}
- K^2 r t^{4/3} \ddot{\mathcal{Q}} + K^3 r t^{2} \mathcal{Q}'' -\frac{1}{3} K^2 r t^{1/3} \dot{\mathcal{Q}} +2K^3 t^{2} \mathcal{Q}' +\frac{5}{3} K^2 r t^{-2/3} \mathcal{Q} -4 r  \mathcal{Q} =0.  
\end{equation}

This wave equation can be solved by the method of separation of variables as discussed above. It gives rise to two differential equations with the separation constant $ -m_1^2 $:
\begin{equation}\label{17d}
\ddot{\mathcal{T}} +\frac{1}{3} \frac{\dot{\mathcal{T}}}{t} +\left( -\frac{5}{3t^{2}} +\frac{4}{K^2 t^{4/3}} + m_1^2 K t^{2/3} \right)\mathcal{T}  =0,
\end{equation}
\begin{equation}\label{17e}
\text{and} \hspace{0.5cm}  {\mathcal{R}''} +  \frac{2\mathcal{R}'}{r} +m_1^2 {\mathcal{R}} =0
\end{equation}
The solutions are respectively:
\begin{equation}\label{17f}
\mathcal{T}(t) = p_1 t^{5/3} {\exp(\xi t^{4/3})} \Sigma  +p_2 t^{5/3} \exp(\xi t^{4/3})
 \Sigma  \left( \int \dfrac{\exp(-2\xi t^{4/3}) dt }{t^{11/3} \Sigma^2} \right) ,
\end{equation}
\begin{equation}\label{17g}
\text{and} \hspace{0.5cm}  \mathcal{R}(r) = \frac{1}{r} \left( p_3 \cos m_1r + p_4 \sin m_1r \right) .
\end{equation}
Here, $ p_1 $, $ p_2 $, $ p_3 $ and $ p_4 $ are the integration constants, \hspace{0.1cm}
$  \xi =\dfrac{3}{4} \sqrt{-K m_{ \textcolor[rgb]{0.00,0.00,1.00}{ 1 } }^2 } $, and
$ \Sigma =\text{HeunB}\left( 4, 0, 0, \dfrac{6\sqrt{6}i}{K^2 (-Km_1^2)^{1/4}}, \dfrac{\sqrt{6}i}{2} (-K m_1^2)^{1/4} t^{2/3} \right) $ represent the Heun function, the solution for Heun's biconfluent equation. The expression for $ \mathcal{Q}(t,r) $ can now be written in terms of $ \mathcal{T}(t) $ and $ \mathcal{R}(r) $. Plugging it in equation \eqref{17a} and using the expression of $ b(t) $ from equation \eqref{5hi}, the expression for $ h_0(t,r) $ is found to be: 
\begin{equation}\label{17h}
h_0(t,r)= K^2 r t^{4/3} \mathcal{Q}(t,r).
\end{equation}

Subsequently, equation \eqref{12c} gives the expression $ h_1(t,r) $ as:
\begin{equation}\label{17i}
h_1(t,r)=  f_1(t)+ \frac{1}{K t^{2/3}} \int_{r_0}^{r}\left( \dot{h}_0(t,r) - \frac{h_0(t,r)}{3t} \right) dr ,
\end{equation}
where $ f_1(t) $ is the integration constant, and $ r_0 $ represents the initial hypersurface from which GWs originate. The arbitrary function $ f_1(t) $ is chosen to be zero, and we find
\begin{equation}\label{17j}
h_1(t,r) = K ( t^{2/3} \dot{\mathcal{T}} +t^{-1/3} \mathcal{T} ) \int_{r_0}^{r}  r \mathcal{R} dr
= K ( t^{2/3} \dot{\mathcal{T}} +t^{-1/3} \mathcal{T} ) \left[  \frac{1}{m_1} \left(p_3 \sin m_1r - p_4 \cos m_1r \right) \right]_{r_0}^{r} .
\end{equation}

\bigskip

\underline{\textbf{Case II: $ \alpha =0 $ and $ \beta =3 $}}

\bigskip

We have from equation \eqref{AG1} :
\begin{equation} \label{18a}
h_0(t,r)= b(t)^3 \mathcal{Q}(t,r).
\end{equation}
Then equation \eqref{AG2} reads:
\begin{equation} \label{18b}
-b^3 \ddot{\mathcal{Q}} +b^4 \mathcal{Q}'' -\frac{5}{2} b^2\dot{b} \dot{\mathcal{Q}} -\frac{5}{2} b^2\ddot{b} \mathcal{Q} +\frac{5}{2} b\dot{b}^2 \mathcal{Q} -b \mathcal{Q} \left\lbrace l(l+1)-2 \right\rbrace =0.
\end{equation}
Using equation \eqref{5hi} and putting $ l=2 $, equation \eqref{18b} gives:
\begin{equation}\label{18c}
-K^3 t^{2} \ddot{\mathcal{Q}} +K^4 t^{8/3} \mathcal{Q}'' -\frac{5}{3} K^3 t \dot{\mathcal{Q}} +\frac{5}{3} K^3  \mathcal{Q} -4K t^{2/3} \mathcal{Q} =0.
\end{equation}

Using the method of separation of variables by the assumption \eqref{AG3}, the wave equation \eqref{18c} leads to two differential equations with the separation constant $ -m_2^2 $:
\begin{equation}\label{18d}
\ddot{\mathcal{T}} +\frac{5}{3} \frac{\dot{\mathcal{T}}}{t} +\left( -\frac{5}{3t^{2}} +\frac{4}{K^2 t^{4/3}} + m_2^2 K t^{2/3} \right)\mathcal{T}  =0,
\end{equation}
\begin{equation}\label{18e}
\text{and} \hspace{0.5cm}  {\mathcal{R}''}  +m_2^2 {\mathcal{R}} =0,
\end{equation}
with the respective solutions:
\begin{equation}\label{18f}
\mathcal{T}(t) = q_1 t \exp(\xi t^{4/3}) \Sigma  +q_2 t \exp(\xi t^{4/3})
 \Sigma  \left( \int \dfrac{\exp(-2\xi t^{4/3}) dt }{t^{11/3} \Sigma^2} \right) ,
\end{equation}
\begin{equation}\label{18g}
\mathcal{R}(r) = q_3 \cos m_2r + q_4 \sin m_2r .
\end{equation}
$ q_1 $, $ q_2 $, $ q_3 $ and $ q_4 $ are the integration constants, \hspace{0.1cm} $ \xi =\dfrac{3}{4} \sqrt{-K m_2^2} $, and
 $ \Sigma =\text{HeunB}\left( 4, 0, 0, \dfrac{6\sqrt{6}i}{K^2 (-Km_2^2)^{1/4}}, \dfrac{\sqrt{6}i}{2} (-K m_2^2)^{1/4} t^{2/3} \right)  $ is the solution for Heun's biconfluent equation. $ \mathcal{Q}(t,r) $ can thus be expressed in terms of $ \mathcal{T}(t) $ and $ \mathcal{R}(r) $. Hence, using equations \eqref{16a} and 
\eqref{5hi}, the expression for $ h_0(t,r) $ is obtained as: 
\begin{equation}\label{18h}
h_0(t,r)= K^3 t^{2} \mathcal{Q}(t,r).
\end{equation}

Finally, $ h_1(t,r) $ is found from equation \eqref{12c}.
\begin{equation}\label{18i}
h_1(t,r)=   f_2(t)+ \frac{1}{K t^{2/3}} \int_{r_0}^{r}\left(   \dot{h}_0(t,r) - \frac{h_0(t,r)}{3t} \right) dr ,
\end{equation}
The arbitrary integration constant $ f_2(t) $ is taken to be zero. 
Hence, we get
\begin{equation}\label{18j}
h_1(t,r) = K^2 \left(t^{4/3} \dot{\mathcal{T}} +\frac{5}{3} t^{1/3} \mathcal{T} \right) \int_{r_0}^{r}  \mathcal{R} dr
= K^2 \left(t^{4/3} \dot{\mathcal{T}} +\frac{5}{3} t^{1/3} \mathcal{T} \right)  \left[\frac{1}{m_2} (q_3 \sin m_2r  -q_4 \cos m_2r) \right]_{r_0}^{r}.
\end{equation}

\bigskip

\underline{\textbf{Case III: $ \alpha =2 $ and $ \beta =0 $}}

\bigskip

From equation \eqref{AG1} we get:
\begin{equation} \label{19a}
h_0(t,r)= r^2 \mathcal{Q}(t,r).
\end{equation}
The equation \eqref{AG2} then becomes:
\begin{equation} \label{19b}
-r^2 \ddot{\mathcal{Q}} +r^2 b \mathcal{Q}'' +\frac{7r^2 \dot{b}}{2b} \dot{\mathcal{Q}} +4rb \mathcal{Q}'
 +\frac{r^2 \ddot{b}}{2b} \mathcal{Q} -\frac{2r^2 \dot{b}^2}{b^2} \mathcal{Q} +2b \mathcal{Q}
 -\frac{r^2}{b^2} \mathcal{Q} \left\lbrace l(l+1)-2 \right\rbrace =0.
\end{equation}
Using equation \eqref{5hi} for $ l=2 $, equation \eqref{19b} reduces to:
\begin{equation}\label{19c}
-r^{2} \ddot{\mathcal{Q}} +K t^{2/3} r^2 \mathcal{Q}'' +\frac{7}{3} \frac{r^2}{t} \dot{\mathcal{Q}} +4Kt^{2/3} r  \mathcal{Q}' -\frac{4r^2}{K^2 t^{4/3}} \mathcal{Q} -\frac{r^2}{t^2} \mathcal{Q} +2K t^{2/3} \mathcal{Q} =0.
\end{equation}
Proceeding as before, this wave equation \eqref{19c} gives rise to two differential equations with the separation constant $ -m_3^2 $:
\begin{equation}\label{19d}
\ddot{\mathcal{T}} -\frac{7}{3} \frac{\dot{\mathcal{T}}}{t} + \left( \frac{1}{t^{2}} +\frac{4}{K^2 t^{4/3}} + m_3^2 K t^{2/3} \right) \mathcal{T}  =0,
\end{equation}
\begin{equation}\label{19e}
\text{and} \hspace{0.5cm}  \mathcal{R}'' +\frac{4 \mathcal{R}'}{r} + \frac{2}{r^2} \mathcal{R} +m_3^2 \mathcal{R} =0,
\end{equation}
The respective solutions are as follows:
\begin{equation}\label{19f}
\mathcal{T}(t) = w_1 t^3 \exp(\xi t^{4/3}) \Sigma  +w_2 t^3 \exp(\xi t^{4/3})
 \Sigma  \left( \int \dfrac{\exp(-2\xi t^{4/3}) dt }{t^{11/3} \Sigma^2} \right) ,
\end{equation}
\begin{equation}\label{19g}
\mathcal{R}(r) = \frac{1}{r^2}(w_3 \cos m_3r + w_4 \sin m_3r) .
\end{equation}
Here, $ w_1 $, $ w_2 $, $ w_3 $ and $ w_4 $ are the integration constants, \hspace{0.1cm} $ \xi =\dfrac{3}{4} \sqrt{-K m_3^2} $, and
 $ \Sigma =\text{HeunB}\left( 4, 0, 0, \dfrac{6\sqrt{6}i}{K^2 (-Km_3^2)^{1/4}}, \dfrac{\sqrt{6}i}{2} (-K m_3^2)^{1/4} t^{2/3} \right)  $. In this case,  
\begin{equation}\label{19h}
h_1(t,r)=   f_3(t)+ \frac{1}{K t^{2/3}} \int_{r_0}^{r}\left( \dot{h}_0(t,r) - \frac{h_0(t,r)}{3t} \right) dr ,
\end{equation}
$ f_3(t) $ is the arbitrary integration constant, and can be chosen to be zero. 
It follows that
\begin{equation}\label{19i}
h_1(t,r) = K^{-1} \left(\frac{1}{t^{2/3}} \dot{\mathcal{T}} +\frac{1}{3t} \mathcal{T} \right) \int_{r_0}^{r}  r^2\mathcal{R} dr
= K^{-1} \left(\frac{1}{t^{2/3}} \dot{\mathcal{T}} +\frac{1}{3t} \mathcal{T} \right)  \left[\frac{1}{m_3} (w_3 \sin m_3r -w_4 \cos m_3r) \right]_{r_0}^{r}.
\end{equation}

\subsection{POLAR GRAVITATIONAL WAVES}

It can be seen from the polar perturbation equations (\eqref{15a}-\eqref{15e} and \eqref{15h}) that even in the vacuum Kantowski-Sachs universe, there exist complicated couplings among the three perturbing terms $ \chi (t,r) $, $ \psi (t,r) $ and $ \zeta (t,r) $. Differentiating equation \eqref{15c} w.r.t. $ r $, we get
\begin{equation}\label{20a}
\dot{\zeta}'= -\frac{\dot{a}}{a}\zeta' -\chi'' +\left(\frac{1}{b^2}-1\right) \psi'',
\end{equation}
which when substituted in equation \eqref{15h} eliminates the $ \dot{\zeta}' $-term and yields
\begin{equation}\label{20b}
\begin{split}
\left\lbrace b\ddot{b} + \frac{\dot{a}^2 b^2}{a^2} - \frac{l(l+1)}{ 4 } (a^2 -1) \right\rbrace \chi
+ \left\lbrace a\ddot{a} + b\ddot{b} + \frac{\dot{a}^2 b^2}{a^2} + \frac{a^2 \dot{b}^2}{b^2} -  \frac{l(l+1)}{ 4 } (a^2 -1) \right\rbrace \psi \qquad\qquad\qquad \\
+ \left\lbrace - \frac{\dot{a} b^2}{a} +\frac{1}{2} b\dot{b} +\frac{1}{2} a\dot{a} b^2 +\frac{1}{2} a^2 b\dot{b} \right\rbrace \dot{\chi}
+ \left\lbrace - \frac{\dot{a} b^2}{a} - \frac{a^2 \dot{b}}{b} + \frac{1}{2} a\dot{a} +\frac{1}{2} b\dot{b} +\frac{1}{2} a\dot{a} b^2 +\frac{1}{2} a^2 b\dot{b} \right\rbrace  \dot\psi \\
 + \frac{1}{2} b^2 \ddot{\chi} - \frac{1}{2} b^2 \chi'' + \frac{1}{2} (a^2+b^2) \ddot{\psi} - \frac{1}{2} (b^2 -1) \psi'' + \left\lbrace \frac{\dot{a} b^2}{a} - b\dot{b} \right\rbrace \zeta' =0.
\end{split}
\end{equation}

Plugging in the expressions for $ a(t) $ and $ b(t) $ from equation \eqref{5hi}, and choosing $ l=2 $, equations  \eqref{15a}-\eqref{15e} and \eqref{20b} are read as:
\begin{equation}\label{20c}
- 3 \zeta + \frac{2 K^2 t^{1/3}}{3}  \chi' + \left(\frac{2 K^2 t^{1/3}}{3} - \frac{1}{3t} \right) \psi' + \dot{\psi}' =0,
\end{equation}

\begin{equation}\label{20d}
\begin{split}
\left( \frac{1}{3t} + \frac{1}{3t^{1/3}} - \frac{2}{3 K^{1/2} t^{4/3}} \right) \chi + \left( \frac{1}{3t} + \frac{1}{3t^{1/3}} - \frac{2}{3 K^{1/2} t^{4/3}} -\frac{4}{3 K^2 t^{7/3}} \right)\psi  
+ K t^{2/3}\dot{\chi}
 \\ + \left( K t^{2/3} + \frac{1}{K t^{4/3}} \right) \dot{\psi}  - K t^{2/3} \zeta'  =0,
\end{split}
\end{equation}

\begin{equation}\label{20e}
- \frac{1}{3t} \zeta + \dot{\zeta} - \chi' + \left(\frac{1}{K^2 t^{4/3}} -1\right) \psi' =0,
\end{equation}

\begin{equation}\label{20f}
\begin{split}
\left( \frac{1}{Kt^{2/3}} - \frac{4K^2}{9t^{2/3}} -3 \right) \chi
+ \left( \frac{1}{Kt^{2/3}} - \frac{4K^2}{9t^{2/3}} - \frac{3}{  K^{3} t^{2} }
+ \frac{4}{9K t^{8/3}} - 3 \right) \psi
 - \frac{2}{3} K^2 t^{1/3} \dot{\chi} \\
 - \left( \frac{2}{3} K^2 t^{1/3}
  -   \frac{1}{3  K t^{5/3}  } + \frac{2}{3 K^2 t^{7/3}} \right) \dot{\psi}
+ \psi'' + \frac{4}{3} K^2 t^{1/3} \zeta'  =0,
\end{split}
\end{equation}

\begin{equation}\label{20g}
 4  \chi
+ \left( - \frac{4}{9t^2}  -  \frac{3}{K^2 t^{4/3}}  + 4  \right) \psi
- \frac{2}{3} K^2 t^{1/3} \dot{\chi} + \left( \frac{2}{3t} - \frac{2}{3} K^2 t^{1/3} \right) \dot{\psi} - \ddot{\psi} =0,
\end{equation}

\begin{equation}\label{20h}
\begin{split}
\left( -\frac{K^2}{9t^{2/3}} - \frac{3}{  2  Kt^{2/3}} + \frac{3}{  2  } \right) \chi
+ \left( - \frac{K^2}{9t^{2/3}}  +  \frac{8}{9K t^{8/3}} - \frac{3}{ 2 Kt^{2/3}} + \frac{3}{  2  } \right) \psi
 + \left( \frac{2K^2 t^{1/3}}{3} + \frac{K}{6t^{1/3}} \right) \dot{\chi} \\
+ \left( \frac{2K^2 t^{1/3}}{3} + \frac{K}{6t^{1/3}} - \frac{5}{6K t^{5/3}} \right) \dot\psi
+ \frac{1}{2} K^2 t^{4/3} \ddot{\chi} + \frac{1}{2} \left( K^2 t^{4/3} + \frac{1}{K t^{2/3}} \right) \ddot{\psi} \\
- \frac{1}{2} K^2 t^{4/3} \chi'' - \frac{1}{2} (K^2 t^{4/3} -1) \psi'' - K^2 t^{1/3} \zeta' =0.
\end{split}
\end{equation}

Equation \eqref{20h} appears to be a wave equation for $ \chi (t,r) $ and $ \psi (t,r) $ with $ \ddot{()} + ()''$ terms. Unlike the case of axial perturbations, the solutions for polar modes cannot be derived easily even in the vacuum case for $ l \geq 2 $. In the polar case, the equation \eqref{15c} involving the three perturbation elements cannot be utilised in simplifying the equations to extract a differential equation in a single variable. However, we may progress if the perturbation due to $ \chi (t,r) $ (or $ \psi (t,r) $) can be neglected, or if $ \chi (t,r) $ and $ \psi (t,r) $ are proportional to each other. Accordingly, we seek analytical solutions for particular cases in the following sequel.

\bigskip

\underline{\textbf{Case I: $ \chi (t,r) $ and $ \psi (t,r) $) are proportional to each other}}

\bigskip

If we consider
\begin{equation}\label{21a}
\psi (t,r) =q \chi (t,r),
\end{equation}
$ q $ being a constant, then equation \eqref{20g} leads us to:
\begin{equation}\label{21b}
\left( 4 + 4q  - \frac{4q}{9t^2}  -  \frac{3q}{K^2 t^{4/3}} \right) \chi + \left( -\frac{2}{3} K^2 t^{1/3} -\frac{2}{3} qK^2 t^{1/3} + \frac{2q}{3t} \right) \dot{\chi} -  q \ddot{\chi} = 0.
\end{equation}
Its solution is given by
\begin{equation}\label{21c}
\chi (t,r) =
F_1(r) t^{  4/3  } \exp \left( \dfrac{  9t^{2/3}  }{K^2} \right) B_1
+ F_2(r) t^{  1/3  } \exp \left( \dfrac{  9t^{2/3}  }{K^2} \right) B_2
= t^{  1/3  }
\exp \left( \dfrac{ 9t^{2/3} }{K^2} \right) \left[  t   F_1(r) B_1  + F_2(r) B_2 \right],
\end{equation}
where
\begin{equation*}
B_1 = \text{HeunB} \hspace{0.1cm} \left( \frac{ 3 }{2} ,
\hspace{0.1cm} \frac{  72  q^2 (1+q)}{(-2q (1+q))^{3/2} K^3},
\hspace{0.1cm} \frac{-(1+q) K^6 - 324 q}{2K^6(1+q)},
\hspace{0.1cm} -\frac{27q}{K^3 \sqrt{-2q (1+q)}}, \hspace{0.1cm} \frac{\sqrt{-2q (1+q)} Kt^{2/3}}{2q} \right) ,
\end{equation*}
\begin{equation}\label{21d}
B_2 = \text{HeunB} \hspace{0.1cm} \left( -\frac{  3 }{2} ,
\hspace{0.1cm} \frac{ 72  q^2 (1+q)}{(-2q (1+q))^{3/2} K^3},
\hspace{0.1cm} \frac{-(1+q) K^6 - 324 q}{2K^6(1+q)},
\hspace{0.1cm} -\frac{27q}{K^3 \sqrt{-2q (1+q)}}, \hspace{0.1cm} \frac{ \sqrt{-2q (1+q)} Kt^{2/3} }{2q} \right).
\end{equation}
Here $ B_1 $ and $ B_2 $ represent two different biconfluent Heun's functions, and $ F_1(r) $ and $ F_2(r) $ are arbitrary functions of $ r $.
Hence, from equation \eqref{21a}, we get
\begin{equation}\label{21e}
\psi(t,r) = q t^{  1/3  }
\exp \left( \dfrac{ 9t^{2/3} }{K^2} \right) \left[  t  F_1(r) B_1  + F_2(r) B_2 \right].
\end{equation}

Moreover, from equation \eqref{20c}, we find that
\begin{equation}\label{21f}
\begin{split}
\zeta (t,r) = \frac{1}{  9  K^2}
\exp \left( \dfrac{  9t^{2/3}  }{K^2} \right)
\left[  \frac{2}{t^{5/3}}  \left\lbrace (1+q) K^4 t^{  10/3  } + 9q t^{ 8/3  }
+  \frac{3}{2}  qK^2  t^2   \right\rbrace  B_1 F'_1
+  2 \left\lbrace  (1+q) K^4 t^{  2/3  } + 9q \right\rbrace B_2 F'_2
\right. \\ \left.
+ K^3 \sqrt{-2q (1+q)} \left\lbrace  t   \tilde{B}_1 F'_1
+ \tilde{B}_2 F'_2  \right\rbrace \right].
\end{split}
\end{equation}
The tilde over `$ B $'- s denotes the $ z $-derivative of the biconfluent Heun's function, $\textrm{HeunB}(\alpha , \beta , \gamma , \delta , z )$.

\bigskip

\underline{\textbf{Case II: $ \chi (t,r) $ and its derivatives are all zero}}

\bigskip

Thus equation \eqref{20g} reduces to:
\begin{equation}\label{22a}
\left(  4  - \frac{4}{9t^2}  -  \frac{3}{K^2 t^{4/3}} \right) \psi + \left( -\frac{2}{3} K^2 t^{1/3} + \frac{2}{3t} \right) \dot{\psi} -  \ddot{\psi} = 0,
\end{equation}
whose solution is
\begin{equation}\label{22b}
\psi (t,r) = t^{  1/3  } \exp \left( -\frac{   t^{2/3} (t^{2/3}K^4 +18)  }{2K^2} \right) [  t  F_3(r)  B_3 + F_4(r) B_4 ] ,
\end{equation}
\begin{eqnarray}\label{22c}
B_3 = \text{HeunB} \hspace{0.1cm} \left( \frac{  3  }{2},
\hspace{0.1cm} \frac{  18  \sqrt{2}}{K^3},
\hspace{0.1cm} \frac{(K^6 +   324   )}{2K^6},
\hspace{0.1cm} -\frac{27 \sqrt{2}}{2K^3}, \hspace{0.1cm} \frac{ \sqrt{2} Kt^{2/3} }{2} \right),   \\
B_4 = \text{HeunB} \hspace{0.1cm} \left( -\frac{ 3  }{2},
\hspace{0.1cm} \frac{ 18  \sqrt{2}}{K^3},
\hspace{0.1cm} \frac{(K^6 +  324  )}{2K^6},
\hspace{0.1cm} -\frac{27 \sqrt{2}}{2K^3}, \hspace{0.1cm} \frac{ \sqrt{2} Kt^{2/3} }{2} \right).
\end{eqnarray}

Plugging this expression in equation \eqref{20c}, we obtain $ \zeta (t,r) $ in the form
\begin{equation}\label{22d}
\begin{split}
\zeta (t,r) = \frac{ 1 }{9K^2 t^{2/3}} \exp \left( -\frac{  t^{2/3} (t^{2/3}K^4 +18)  }{2K^2} \right)
\left[  \left\lbrace  3  K^2  t
-  18 t^{5/3}  \right\rbrace B_3 F'_3 -  18 t^{2/3}  B_4 F'_4
+ \sqrt{2}  K^3 t^{2/3}  \left\lbrace  t  \tilde{B}_3 F'_3 + \tilde{B}_4 F'_4 \right\rbrace   \right] ,
\end{split}
\end{equation}
where $ F_3(r) $ and $ F_4(r) $ are arbitrary functions of $ r $. In this case also, we see that two different biconfluent Heun's functions represented by $ B_3 $ and $ B_4 $ appear in the $ t $-solution. The tilde over `$ B $'- s  denotes the $ z $-derivative of $\textrm{HeunB}(\alpha , \beta , \gamma , \delta , z )$.

\bigskip

\underline{\textbf{Case III: $ \psi (t,r) $ and its derivatives are all zero}}

\bigskip

In this case, equation \eqref{20g} reads as:
\begin{equation}\label{23a}
 4  \chi - \frac{2}{3} K^2 t^{1/3} \dot{\chi} = 0,
\end{equation}
giving the solution
\begin{equation}\label{23b}
\chi (t,r) = F_5(r) \exp \left[\frac{9t^{2/3}}{K^2}\right].
\end{equation}
Here $ F_5 (r) $ is arbitrary. Thus equation \eqref{20c} yields
\begin{equation}\label{23c}
\zeta (t,r) = \frac{2}{9} K^2 t^{1/3} \exp \left[\frac{9t^{2/3}}{K^2}\right] F'_5.
\end{equation}

Now, we know that the $ t $ and $ r $-solutions separate out as product in the Regge-Wheeler formalism, so one can express the polar perturbation terms as:
\begin{equation}\label{24a}
\nu (t,r) = \mathcal{T}_{\nu} (t) \mathcal{R}_{\nu} (r), \hspace{1cm} \nu =\chi, \psi, \zeta,
\end{equation}
analogous to the assumption \eqref{AG3} used in the separation of variables for axial solutions. In all the three cases we have analysed for polar modes, the radial part of the solutions for the perturbation parameters remains arbitrary. However, the set of equations \eqref{20c}-\eqref{20h} shows on inspection that
\begin{equation}\label{24b}
\zeta (t,r) \propto \mathcal{R}_{\mu}'(r), \hspace{0.6cm} \text{or} \hspace{0.6cm} \zeta (t,r) \propto \int \mathcal{R}_{\mu}(r) dr ,
\end{equation}
where $ \mu = {\chi, \psi } $ and $ \mathcal{R}(r) $ is the radial solution for the perturbations $ \chi (t,r) $ and $ \psi (t,r) $.  It can be inferred from equation \eqref{24b} that $ \mathcal{R}_{\mu} (r) $ will have sinusoidal nature. So we may write:
\begin{equation}\label{24c}
\mathcal{R}_{\mu} (r) = \lambda_1 \sin (Mr) + \lambda_2 \cos (Mr),
\end{equation} and hence
\begin{equation}\label{24d}
| \mathcal{R}_{\zeta} (r)| = \lambda_3 \left[\lambda_1 \cos (Mr) - \lambda_2 \sin (Mr)\right].
\end{equation}
Here, $ \lambda_1 $, $ \lambda_2 $, $ \lambda_3 $ and $ M $ are all constants.

\section{Constraining the perturbation solutions from observational data}
\begin{figure}
\centering
\includegraphics[scale=0.48]{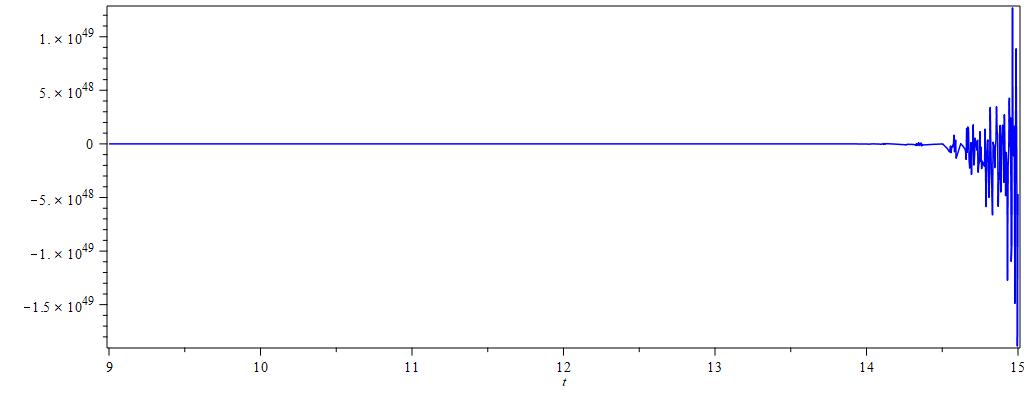}
\caption{Graph showing the variation of $ \mathcal{T_{\mu}} (t) $ with time $t$ in the interval 9 to 15 (on an arbitrary scale).}
\label{fig:Polar1}
\end{figure}

\begin{figure}
\centering
\includegraphics[scale=0.48]{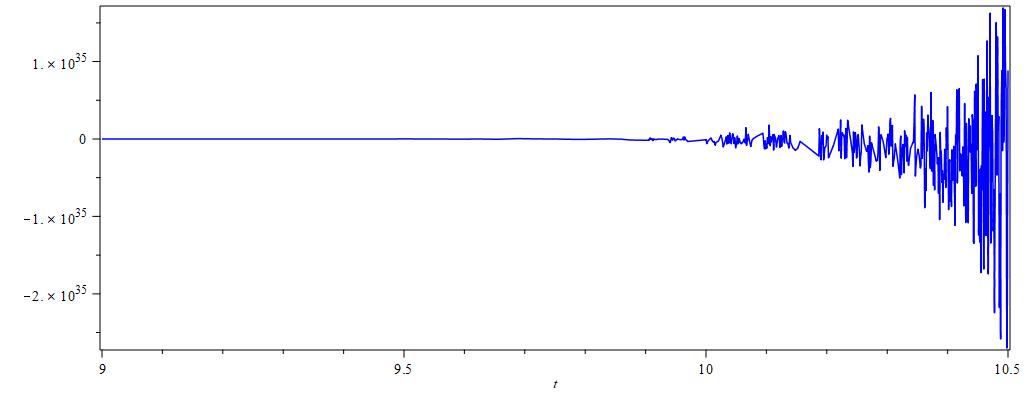}
\caption{Graph showing the variation of $ \mathcal{T_{\mu}} (t) $ with time $t$ for a different interval of time.}
\label{fig:Polar2}
\end{figure}

\begin{figure}
\centering
\includegraphics[scale=0.48]{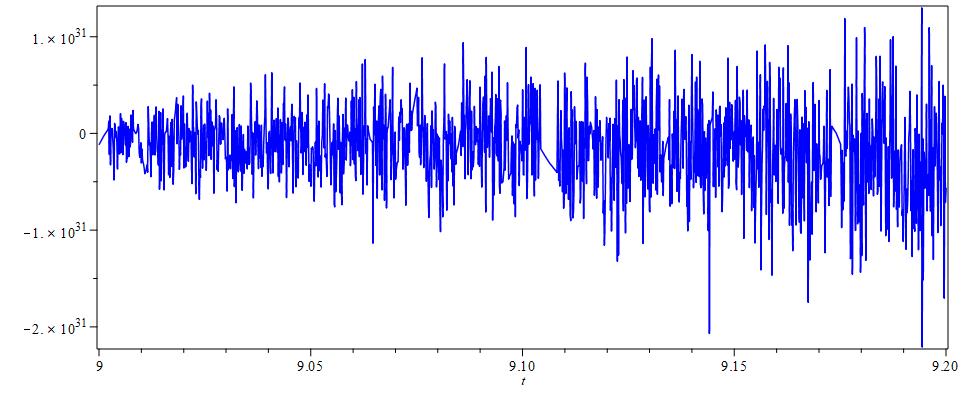}
\caption{Graph showing the variation of $ \mathcal{T_{\mu}} (t) $ with time $t$ for a smaller interval of time to get a clearer estimate of the frequency.}
\label{fig:Polar3}
\end{figure}

Aiming to place constraints on the parameters appearing in the solutions, we compare our theoretical results with observational data available in the literature. Here we concentrate on  the polar solutions, with particular emphasis on Case I. Without any loss of generality we may assume that $ F_1(r) = F_2(r) = \mathcal{R_{\mu}}(r) $, because all three are arbitrary functions of $r$, and choosing $ q= K=1 $, equations \eqref{21c} and \eqref{21e} reduce to
\begin{equation}\label{25a}
\chi (t,r)= \psi (t,r)
= t^{1/3} \exp \left(  9t^{2/3} \right) \left( t B_1  +  B_2 \right) \mathcal{R_{\mu}}(r).
\end{equation}
The temporal part of this solution, given by $  \mathcal{T_{\mu}} (t) = t^{1/3} \exp \left( 9t^{2/3} \right) (t B_1 + B_2), $  is plotted as a function of time  $ t $. The unit of time is chosen arbitrarily, varying only in the scale and not in dimension. However the time in the argument of $\mathcal{T_{\mu}}$ must be on the same scale. Sample plots are given in FIG.~\ref{fig:Polar1}, FIG.~\ref{fig:Polar2}, and FIG.~\ref{fig:Polar3}, which are obtained by substituting $ q= K=1 $ in both the Heun's functions, namely, $B_1$ and $B_2$. As the solutions are exponentially increasing with time, the amplitudes keep increasing with time. For the sake of simplicity we have illustrated the amplitude variation only over a small range of time (chosen on an arbitrary scale).

Analysing these plots we obtain an order of magnitude estimate of the frequency to be around 1500 Hz, so that we may say that the frequency of polar GWs in Kantowski-Sachs universe lies in the probable range 1000-2000 Hz. This estimate may be understood by examining the figures closely, especially FIG.~\ref{fig:Polar3}. From the solution \eqref{22b} for $ \psi (t,r) $ in Case II, we find that the temporal nature of the perturbation will be similar to that in Case I, the only difference arising from the exponential factor. In Case III, the solutions are more of theoretical interest. The perturbation $ \chi (t,r) $ shows an exponential growth with time (equation \eqref{23b}). Similar is the behaviour of $ \zeta (t,r) $ w.r.t. time (equation \eqref{23c}). In all three cases, the radial solutions remain arbitrary and may be assumed to be represented by \eqref{24c} and \eqref{24d} as mentioned earlier.

Corresponding to the frequency range 1000-2000 Hz, the available observational data is of the GW strain \textit{h}, which is of the order of $ 10^{-24} $ \cite{PAPA1, PAPA2}. This value is used to constrain the parameters appearing in the expression for the strain obtained from our theoretical analysis.

Let us take two nearby particles located at spacetime points $ x_1^{\alpha} = (0,0,0,0) $ and $ x_2^{\alpha} = (0,\epsilon_r,0,0) $ along the radial direction, both initially at rest. The proper distance between the particles can be determined from the following relation \cite{SCHUTZ} :
\begin{equation}\label{25b}
l = \int |ds^2|^{1/2} = \int |g_{\alpha \beta} dx^{\alpha} dx^{\beta}|^{1/2} .
\end{equation}
Since the particles are separated along the radial direction, the integration can be carried over $ dr $. In the unperturbed KS spacetime \eqref{1}, the proper distance between them is found to be
\begin{equation}
L^{(0)} =  \int_0^{\epsilon_r} |g_{r r}^{(0)} dr^2|^{1/2} = \int_0^{\epsilon_r} | -a(t)^2 |^{1/2} dr = a(t) \epsilon_r.
\end{equation}
When the background spacetime is subjected to the polar modes of gravitational perturbations as described by the line element \eqref{6d}, this distance becomes:
\begin{eqnarray}\label{25c}
L & = & \int_0^{\epsilon_r} |g_{r r} dr^2|^{1/2}
\simeq \int_0^{\epsilon_r} | -a(t)^2 + eY(\theta) \left\lbrace \chi (t,r) + \psi (t,r) \right\rbrace |^{1/2} dr \nonumber  \\
& & \simeq \int_0^{\epsilon_r} | -a(t)^2 |^{1/2} \left[1 - \frac{1}{2a(t)^2} \cdot eY(\theta) \left\lbrace \chi (t,r) + \psi (t,r) \right\rbrace \right] dr.
\end{eqnarray}
It is to be noted that the polar perturbation matrix \eqref{6c} in RW gauge exhibits off-diagonal terms ($ g_{tr} $ and $ g_{rt} $) involving $ \zeta (t,r) $. In deriving the proper distance, its contribution can be neglected in comparison to the diagonal term $ g_{rr} $ containing $ a(t)^2 $, $ \chi (t,r) $ and $ \psi (t,r) $. Now, collecting the expressions from equations \eqref{21a}, \eqref{21c}, \eqref{21e} and \eqref{24c}, we have
\begin{equation}\label{25d}
\chi (t,r) + \psi (t,r) = (1+q) \mathcal{T}_{\mu}(t) \mathcal{R}_{\mu}(r)
= (1+q)  \mathcal{T}_{\mu}(t) \left[ \lambda_1 \sin (Mr) + \lambda_2 \cos (Mr) \right].
\end{equation}
Therefore
\begin{eqnarray}\label{25e}
 L & \simeq & \hspace{0.2cm}  a(t) \epsilon_r - \frac{e (1+q) Y(\theta)}{2a(t)} \mathcal{T}_{\mu}(t) \int_0^{\epsilon_r} \left[ \lambda_1 \sin (Mr) + \lambda_2 \cos (Mr) \right] \nonumber \\
& = & a(t) \epsilon_r - \frac{e}{M a(t)} Y(\theta) \mathcal{T}_{\mu}(t)
\left[ \lambda_1 \left\lbrace 1- \cos (M \epsilon_r) \right\rbrace + \lambda_2 \sin (M \epsilon_r) \right].
\end{eqnarray}
We have inserted $ q=1 $ to arrive at the last expression. Finally, the strain can be obtained as
\begin{equation}\label{25f}
h = \frac{\Delta L}{L} = \frac{L- L^{(0)}}{L^{(0)}} \simeq \frac{eY(\theta)}{M \epsilon_r a(t)^2 } \mathcal{T}_{\mu}(t)
\left[ \lambda_1 \left\lbrace \cos (M \epsilon_r) -1 \right\rbrace - \lambda_2 \sin (M \epsilon_r) \right].
\end{equation}
We know that, $ Y_{20} (\theta) = \dfrac{1}{4} \sqrt{\dfrac{5}{\pi}} (3 \cos^2 \theta  -1) $, and moreover, equation \eqref{5hi} gives $ a(t)^2 = t^{-2/3} $, with $ K=1 $. Putting $ Y_{20} = 0.158 $ for $ \theta = 45^{\circ} $, and inserting the expression for $ a(t) $ in terms of $ t $, we get from equation \eqref{25f} :
\begin{equation}\label{25g}
h \simeq \left( \frac{0.158 e}{M \epsilon_r} \right) t \exp \left( 9t^{2/3} \right) (t B_1 + B_2)
\left[ \lambda_1 \left\lbrace \cos (M \epsilon_r) -1 \right\rbrace - \lambda_2 \sin (M \epsilon_r) \right].
\end{equation}
To remain in agreement with the available data, this quantity is constrained to be of the order of $ 10^{-24} $. However, that requires the following conditions to be fulfilled: (i) $e \ll \epsilon_r$, (ii) $M \epsilon_r $ to be extremely large, and (iii) the parameters $ \lambda_1 $ and $\lambda_2$ to be small enough.

\bigskip

\section{Discussions and conclusions}

In this paper, we have employed the Regge-Wheeler gauge to study the axial as well as polar perturbations on the Kantowski-Sachs background. Starting from the set of linearised Einstein equations for the axially perturbed background \eqref{6b}, we have derived a wave equation in terms of $ \mathcal{Q}(t,r) $ (the equation \eqref{AG2}). Subsequently, $ h_0(t,r) $ and $ h_1(t,r) $ can be evaluated easily (in view of equations \eqref{AG1} and \eqref{12c}). Combining these $ t $ and $ r $-solutions, we arrive at the complete expressions for the axial perturbations. The perturbations are found to be in the form of the product of four functions, each being a function of only one of the coordinates $ t $, $ r $, $ \theta $ and $ \phi $. The $ \theta $-dependence is represented by the term $ \sin \theta (\partial_{\theta}Y) $ in the expressions \eqref{6a}. The $ \phi $-dependence is removed at the beginning by setting the projection of the angular momentum on the $ z $-axis to zero (i.e. $ m = 0 $) in accordance with \cite{RW}.

The pre-factor of $ \mathcal{Q} $ in the wave equation (equations \eqref{AG2}, \eqref{17c}, \eqref{18c}, \eqref{19c}) acts as an effective potential (akin to equations (4.7a) and (4.7b) of \cite{VISH}, equations (87) and (91) of Ref. \cite{REZ}, or equations (24) and (25) of Ref. \cite{RW}). Therefore, the RW equation resembles the wave equation in a scattering potential, as mentioned by Rezzolla \cite{REZ}.
The wave equation \eqref{AG2} contains the first-order derivatives of $ \mathcal{Q} $, in addition to its second-order derivatives. The terms $ \dot{\mathcal{Q}} $ and $ \mathcal{Q}' $ are responsible for the damping of the axial GW as it spreads out in the $ (t-r) $ hypersurface. Previous studies have shown that damping terms appear in LTB background (Eq. (50) of \cite{CL1}) but not in FLRW universe \cite{M1, SH1}. We can therefore conclude that this damping arises from the anisotropy of the background Kantowski-Sachs spacetime. Our previous work \cite{GD} has shown that damping happens in Bianchi I spacetime too.

We observe that the term $ \mathcal{Q}' $ (the first-order $ r $-derivative of $ \mathcal{Q} $) appears in cases where $ r $ is raised to the power $ \alpha \neq 0 $ . The radial solution slightly changes with the presence of $ \mathcal{Q}' $. But the $ r $-dependence of $ h_0(t,r) $ and $ h_1(t,r) $ has the same sinusoidal nature in all the cases.
For all values of $ \alpha $ and $ \beta $, the temporal solution comes out in terms of the same biconfluent Heun’s function \cite{HOR, RON}. But different values of $ \beta $ in the power of $ t $ (appearing as a product term in the temporal solution) will bring about variation in the nature of the axial perturbations $ h_0(t,r) $ and $ h_1(t,r) $.

The approach used in this paper is the same as the one we adopted in the case of Bianchi I background \cite{GD}. Comparing the corresponding results, we find that the nature of temporal and radial solutions of the axial perturbation equations are much alike in both backgrounds. There is a small difference only in the Heun's function. It arises from the difference in the term containing $ l $ in the wave equation. While solving for the axial waves in Bianchi I spacetime, we assumed that $ \cot \theta =1/ \theta $, which holds for small values of $ \theta $. However, this assumption is not required in the present calculations. This is because of the nature of the line element in the Kantowski-Sachs universe.

In \cite{BR1}, the perturbations to Kantowski-Sachs model were analysed using 1+1+2 covariant decomposition of spacetime. It was shown that the perturbations represented by odd and even projections of Weyl tensor obey wave equations. At the geometrical optics approximation, the gravitational modes decouple and propagate with the speed of light, whereas at higher orders, the waves exhibit dispersive behaviour. Our study employing a different approach also yields wave equations for axial perturbations and indicates the damping of the GWs in the anisotropic KS background.

We have chosen $ l=2 $ in our calculations for the wavelike solutions, just as in our previous paper \cite{GD}. It is known that the spherical harmonics with $ l = 0 $ correspond to spherical symmetry, whereas higher values of $ l $ indicate deviation from spherical symmetry, and are associated with non-zero quadrupole moment (which are characteristic features of gravitational radiations). According to Clarkson \textit{et al.} \cite{CL1}, scalars on $ S^2 $ can be expressed as a sum over polar modes, and higher-rank tensors as sums over both the polar and axial modes. A dipole term $ (l = 1) $ will appear in the expansion of scalars and vectors in terms of spherical harmonics, but higher multipoles $ (l \geq 2) $ can be present in tensors of any rank.
We need to consider $ l \geq 2 $ in order to take into account the axial modes coming from the expansion of both vector and tensor functions. The value of $ l $ determines the height of the effective potential barrier, given by the coefficient of $ \mathcal{Q} $ in the master equation for axial modes \eqref{AG2} \cite{RW}. A change in the value of $ l $ in our calculations produces only a small change in the Heun's function appearing in the temporal solution, without affecting the radial solution. Consequently, the expressions for $ h_0(t,r) $ and $ h_1(t,r) $ will be slightly different. Although there are no axial perturbations with $ l = 0 $ \cite{CL1}, but polar perturbations exist for all values: $ l=0,1,2 \cdots $. In the $ (0-0) $ element of the polar perturbation matrix, an additional term $ \eta (t,r) $ appears along with $ \chi (t,r) $ and $ \psi (t,r) $ (See equation \eqref{6e}). We have considered $ \eta =0 $ for the vacuum solutions to avoid cumbersome calculations which are difficult to handle. This assumption by itself acts as a constraint equation \cite{CL1}. In the FLRW case, $ \eta $ has been found to vanish \cite{M2, ROST2}. When $ l=0 $ or 1, i.e. for large angle fluctuations, the field equations do not yield $ \eta =0 $ \cite{CL1}. The perturbed KS metric with $ \eta \neq 0 $ is discussed briefly in Appendix A. The complicated equations arising in this case have been somewhat simplified by specialising to the stiff fluid equation of state. We find that the scale factors of the KS background in vacuum and with stiff fluid are given by similar expressions in the solutions of the field equations.

Our investigations reveal that the polar perturbations (even with $ \eta =0 $) involve far more complicated couplings among the perturbing elements than the axial perturbations to the Kantowski-Sachs background and also their FLRW counterparts. Unlike in the case of FLRW background \cite{M2, SH2, ROST2}, no polar perturbation equation in a single variable can be derived in our case. This poses difficulty in finding analytical solutions for the polar modes. We resort to the vacuum case since absence of matter simplifies the equation set to some extent. Following Clarkson \textit{et al.} \cite{CL1}, we try to solve the perturbation equations analytically in some particular cases, such as assuming $ \chi (t,r) $ and $ \psi (t,r) $ proportional to each other, or neglecting the contribution of either of $ \chi (t,r) $ and $ \psi (t,r) $ in the equations. The solutions for the polar modes may be written as a product of a radial function (equations\eqref{24c}, \eqref{24d}) which is sinusoidal in nature, and a temporal function which is again a combination of biconfluent Heun's functions and their derivatives (equations \eqref{21c}-\eqref{21f}, \eqref{22b}-\eqref{22d}) or contains an exponential function (equations \eqref{23b}, \eqref{23c}).  From the plots for the temporal part of the polar perturbation solutions (FIG.~\ref{fig:Polar1}, FIG.~\ref{fig:Polar2} and FIG.~\ref{fig:Polar3}), we have estimated the frequency range of polar GWs to be 1000-2000 Hz. This lies in the audio frequency range and can be detected by ground-based interferometers. Using the available data for the wave strain corresponding to this range, we have found the constraints on the parameters appearing in the perturbed wave solutions. The same could not be done for the axial waves owing to the complicated nature of the corresponding solutions (due to the second term in \eqref{AG6}).

We find that the azimuthal velocity of the fluid present in the spacetime is perturbed by the propagation of axial GWs and not by the polar GWs. The reverse happens for the remaining components of the fluid velocity: the polar waves cause a change in them, but the axial waves do not influence them. Moreover, the axial GWs leave the matter field unperturbed, whereas the polar waves are responsible for perturbations in the energy density and pressure. It has  been pointed out that polar modes cannot exist if the matter perturbations vanish, that is to say that the polar GWs must trigger matter inhomogeneities and anisotropies in the cosmological background. These properties have been reported in the studies on FLRW background \cite{M2, SH1, SH2, SIDD}. However, at present we have not been able to solve the perturbation equations in presence of matter analytically, and hence cannot comment on the cosmological rotation of the fluid induced by the propagating axial GWs.

We also find that the polar perturbation equations in Kantowski-Sachs background are quite similar to those in Bianchi I background (We are working on polar waves in Bianchi I background, and will report it separately). We intend to explore the nature of GWs in other spacetimes and in modified theories of gravity.

\section*{Appendix A: Polar perturbations with $ \eta \neq 0 $}

It has been mentioned that $ \eta \neq 0 $ for the polar perturbations with $ l=0,1 $ \cite{GMG, CL1}. In this section we derive the linearised field equations for the KS metric with polar perturbations \eqref{6f} in presence of matter. 
The expressions for the perturbed energy-momentum tensor are available from equations \eqref{9b}-\eqref{9g}.
The (0-0) component in this case is different: $T_{tt}= \rho_0 [1+ e(\Delta +\chi+\psi -2\eta) Y] $. The perturbation equations (\eqref{14a}-\eqref{14e} when $ \eta $ is non-zero) are listed below.

\begin{equation}\label{26a}
w= \frac{1}{ab^2 (\rho_0+p_0)} \left[ - \dfrac{l(l+1)}{2} \zeta + b\dot{b} \chi' +\left(-\frac{\dot{a}}{a} - \frac{\dot{b}}{b} + b\dot{b}\right) \psi' -2b \dot{b} \eta' + \dot{\psi}' \right],
\end{equation}

\begin{equation}\label{26b}
\begin{split}
v= \frac{1}{2(\rho_0+p_0)} \left[ \left( -\frac{\dot{a}}{a^3} + \frac{\dot{a}}{a} - \frac{\dot{b}}{a^2b} + \frac{\dot{b}}{b} \right)\chi
+\left( -\frac{\dot{a}}{a^3} + \frac{\dot{a}}{a} - \frac{\dot{b}}{a^2b} + \frac{\dot{b}}{b} - \frac{2\dot{b}}{b^3}\right)\psi
\right. \\ \left.
- 2\left( \frac{\dot{a}}{a} +   \frac{\dot{b}}{b}\right) \eta + \frac{1}{a^2}\dot{\chi} + \left( \frac{1}{a^2}+\frac{1}{b^2}\right) \dot{\psi} - \frac{1}{a^2}\zeta' \right],
\end{split}
\end{equation}

\begin{equation}\label{26c}
u =0,
\end{equation}

\begin{equation}\label{26d}
\frac{\dot{a}}{a}\zeta +\dot{\zeta} -\chi' +\left(\frac{1}{b^2}-1\right) \psi' +2\eta' =0,
\end{equation}

\begin{equation}\label{26e}
\begin{split}
\Delta = \frac{1}{a^4b^6\rho_0} \left[ \left\lbrace -a^4b^4\dot{b}^2 - 2a^3\dot{a}b^5\dot{b} + 2a\dot{a}b^5\dot{b}
-\frac{l(l+1)}{2} \cdot a^2b^4 \right\rbrace \chi + \left\lbrace a^4b^2 - a^4b^4\dot{b}^2
\right. \right. \\ \left. \left.
 + 2a^4b^2\dot{b}^2 -2a^3\dot{a}b^5\dot{b}
+ 2a\dot{a}b^5\dot{b} + 2a^3\dot{a}b^3\dot{b} -\frac{l(l+1)}{2} \cdot a^2b^2(a^2+b^2) \right\rbrace \psi
\right. \\ \left.
+ 2\left( a^4b^4 \dot{b}^2 + 2a^3\dot{a} b^5 \dot{b} \right) \eta - a^2b^5\dot{b}\dot{\chi} - \left( a^2b^5\dot{b} + a^4b^3\dot{b} + a^3\dot{a}b^4 \right) \dot{\psi}
+ a^2b^4\psi'' +2a^2b^5\dot{b}\zeta'  \right],
\end{split}
\end{equation}

\begin{equation}\label{26f}
\begin{split}
\Pi = \frac{1}{a^6b^6p_0} \left[ \left\lbrace a^6b^4\dot{b}^2 + 2a^6b^5\ddot{b} - \frac{l(l+1)}{2} \cdot a^6b^4 \right\rbrace \chi + \left\lbrace -a^6b^2 + a^6b^4\dot{b}^2 - 2a^6b^3\ddot{b}
\right. \right. \\ \left. \left.
+ 2a^6b^5\ddot{b} - \frac{l(l+1)}{2} \cdot a^6b^2(b^2 -1)\right\rbrace \psi - 2 \left\lbrace a^6b^4 \dot{b}^2 + 2 a^6b^5 \ddot{b} - \frac{l(l+1)}{2} \cdot a^6b^4  \right\rbrace \eta
\right. \\ \left.
 + a^6b^5\dot{b}\dot{\chi} + (a^6b^5\dot{b} - a^6b^3\dot{b})\dot{\psi} - 2a^6b^5\dot{b}\dot{\eta} + a^6b^4\ddot{\psi}
 \right],
\end{split}
\end{equation}

\begin{equation}\label{26i}
\begin{split}
p_0 b^2\Pi = \frac{1}{a^4b^2} \left[ (-a\ddot{a}b^4 + a^4b^3\ddot{b} + a^3\ddot{a}b^4 + a^3\dot{a}b^3\dot{b} - a\dot{a}b^3\dot{b} + \dot{a}^2b^4)(\chi+\psi) + (a^4\dot{b}^2 - a^4 b\ddot{b} -  a^3\dot{a}b\dot{b})\psi
\right. \\ \left.
- 2(a^4b^3\ddot{b} + a^3\ddot{a}b^4 + a^3\dot{a}b^3\dot{b}) \eta
+ (-a\dot{a}b^4 + \frac{1}{2}a^3\dot{a}b^4 + \frac{1}{2}a^4b^3\dot{b} + \frac{1}{2}a^2b^3\dot{b})(\dot{\chi}+\dot{\psi})
+ ( - a^4b\dot{b} + \frac{1}{2}a^3\dot{a}b^2)\dot{\psi}
\right. \\ \left.
- (a^4b^3 \dot{b} + a^3 \dot{a} b^4) \dot{\eta} + \frac{1}{2} a^2b^4(\ddot{\chi}+\ddot{\psi}+\chi''+\psi'')
+ \frac{1}{2}a^4b^2 \ddot{\psi} - \frac{1}{2}a^2b^2 \psi'' - a^2b^4 \eta'' - a^2b^3\dot{b}\zeta' - a^2b^4\dot{\zeta}'
\right. \\ \left.
- \frac{l(l+1)}{4}  \left\lbrace a^2b^2(a^2 -1)(\chi+\psi) -
\textcolor[rgb]{0.00,0.00,1.00}{ 2 } a^4b^2 \eta \right\rbrace  \right].
\end{split}
\end{equation}

Inserting equation \eqref{26f} to remove $\Pi$, equation \eqref{26i} reads as
\begin{equation}\label{26j}
\begin{split}
\left[ -a\ddot{a}b^4 - a^4b^3\ddot{b} + a^3\ddot{a}b^4 + a^3\dot{a}b^3\dot{b} - a\dot{a}b^3\dot{b} + \dot{a}^2b^4 -a^4b^2\dot{b}^2
+ \frac{l(l+1)}{4} (a^4b^2 +a^2b^2) \right] \chi \\
+ \left[ -a\ddot{a}b^4 - a^4b^3\ddot{b} + a^3\ddot{a}b^4 + a^3\dot{a}b^3\dot{b} - a\dot{a}b^3\dot{b} - a^3\dot{a}b\dot{b} + \dot{a}^2b^4 + a^4\dot{b}^2 + a^4 b\ddot{b} - a^4b^2\dot{b}^2 +a^4
\right. \\ \left.
+ \frac{l(l+1)}{4} (a^4b^2 +a^2b^2 -2a^4) \right]  \psi
  +   \left[ 2a^4b^3\ddot{b}
 - 2a^3\ddot{a}b^4  - 2a^3\dot{a}b^3\dot{b} + 2a^4b^2  \dot{b}^2
- \frac{l(l+1)}{ 2 } \cdot   a^4b^2  \right] \eta \\
+ \left[ -a\dot{a}b^4 + \frac{1}{2}a^3\dot{a}b^4 - \frac{1}{2}a^4b^3\dot{b} + \frac{1}{2}a^2b^3\dot{b} \right] \dot{\chi}  \\
+ \left[ -a\dot{a}b^4 + \frac{1}{2}a^3\dot{a}b^4 - \frac{1}{2}a^4b^3\dot{b} + \frac{1}{2}a^2b^3\dot{b} + \frac{1}{2}a^3\dot{a}b^2 - a^4b^3\dot{b} \right] \dot{\psi} + (a^4b^3 \dot{b} - a^3 \dot{a} b^4) \dot{\eta}  - a^2b^3\dot{b}\zeta'  \\
 + \frac{1}{2} a^2b^4 \left(\ddot{\chi}+\chi''\right)
+ \left(\frac{1}{2}a^2b^4 - \frac{1}{2}a^4b^2\right) \ddot{\psi} + \left(\frac{1}{2}a^2b^4 - \frac{1}{2}a^2b^2\right) \psi'' - a^2b^4 \eta''
 - a^2b^4\dot{\zeta}' =0 .
\end{split}
\end{equation}

The time-derivatives of $ w $, $ v $ and $ \Delta $ determined hereafter yield unwieldy expressions. These can be simplified to some extent if we consider the matter distribution to be represented by a stiff fluid.

\bigskip

\underline{\textbf{Special case: The stiff fluid}}

\bigskip

For a perfect fluid, the extreme relativistic limit is given by the stiff fluid equation of state:
\begin{equation} \label{27a}
p_0 = \rho_0 .
\end{equation}
Under this condition, the speed of sound equals the speed of light. The field equations for the stiff fluid-filled KS universe are:

\begin{equation} \label{27b}
\frac{2 \dot{a}\dot{b}}{ab} + \frac{\dot{b}^2}{b^2} + \frac{1}{b^2} =\rho_0, \hspace{1cm}
-\left( \frac{2\ddot{b}}{b} +\frac{{\dot{b}}^2}{b^2} + \frac{1}{b^2} \right) = \rho_0, \hspace{1cm}
-\left( \frac{\ddot{a}}{a} +\frac{\ddot{b}}{b} +\frac{\dot{a} \dot{b}}{ab}\right) = \rho_0.
\end{equation}

The relation \eqref{4b}, i.e. $ a=b^n $ is used again to solve the background field equations \eqref{27b}. Proceeding in the same way as was done to find the vacuum solutions in Sec. II, we obtain the same solutions for the scale factors as before, i.e.,
\begin{equation}\label{28a}
b = [(n+2)(k_1 t)]^{\frac{1}{n+2}}, \hspace{0.5cm} \text{and} \hspace{0.5cm}  a = [(n+2)(k_1 t)]^{\frac{n}{n+2}}.
\end{equation}
However, the numerical value of $ n $ remains arbitrary here. We know that its value cannot be taken 0, 1 or -1/2. Let us choose
\begin{equation} \label{28b}
 n=1/2.
\end{equation}


Inserting the conditions \eqref{27a} and \eqref{28b} in equation \eqref{E1}, the continuity equation becomes
\begin{equation} \label{E2}
\dot{\rho_0} + \frac{5\dot{b}}{b} \rho_0 =0 .
\end{equation}

The polar-perturbed equations \eqref{26a}-\eqref{26e} in this particular case are as follows :

\begin{equation}\label{29a}
\begin{split}
w= \frac{1}{2 \rho_0 b^{7/2}} \left[ - \frac{l(l+1)}{2} b \zeta
+ b^{2} \dot{b} \chi' + \left( b^{2} \dot{b} - \frac{3\dot{b}}{2} \right) \psi'
- 2b^{2} \dot{b} \eta' + b \dot{\psi}'  \right],
\end{split}
\end{equation}

\begin{equation}\label{29b}
\begin{split}
v= \frac{1}{8 \rho_0 b^{3}} \left[ \left( 3b^2 \dot{b} - 3b \dot{b} \right) \chi
+ \left( 3b^2 \dot{b} - 3b \dot{b} - 4\dot{b} \right) \psi
- 6{b}^2 \dot{b} \eta + 2b^2 \dot{\chi} + (2b^2 + 2b) \dot{\psi} - 2b^2 \zeta'
\right],
\end{split}
\end{equation}

\begin{equation}\label{29c}
u =0,
\end{equation}

\begin{equation}\label{29d}
\frac{\dot{b}}{2b}\zeta +\dot{\zeta} -\chi' +\left(\frac{1}{b^2}-1\right) \psi' +2\eta' =0,
\end{equation}

\begin{equation}\label{29e}
\begin{split}
\Delta = -\frac{1}{\rho_0 b^{4}} \left[ \left\lbrace 2{b^2} \dot{b}^2 - b\dot{b}^2
+ \frac{l(l+1)}{  2  }b \right\rbrace \chi
+ \left\lbrace 2{b^2} \dot{b}^2 - b \dot{b}^2 - 3\dot{b}^2 + \frac{l(l+1)}{2} \left( b+1 \right) -1 \right\rbrace \psi - 4{b^2}\dot{b}^2 \eta
\right. \\ \left.
 + {b^2} \dot{b} \dot{\chi} + \left\lbrace {b^2} \dot{b} + \frac{3}{2}b \dot{b} \right\rbrace \dot{\psi}
- b \psi'' - 2{b^2} \dot{b} \zeta'  \right] .
\end{split}
\end{equation}

Since $ p_0 = \rho_0 $, their respective perturbations $ \Delta $ and $ \Pi $ are equated. Hence, using equation \eqref{28b} in equations \eqref{26f} and \eqref{26j}, replacing $ \Pi $ by $ \Delta $ and plugging in the expression for $ \Delta $ from equation \eqref{29e}, we get respectively:
\begin{equation}\label{29f}
\begin{split}
-\frac{1}{\rho_0 b^4} \left[ \left\lbrace 2b^3 \ddot{b} + 3b^2 \dot{b}^2 - b \dot{b}^2  - \frac{l(l+1)}{2} (b^2 -b) \right\rbrace \chi
+ \left\lbrace 2b^3 \ddot{b}  - 2b \ddot{b} + 3b^2 \dot{b}^2
\right. \right. \\ \left. \left.
- b \dot{b}^2 - 3\dot{b}^2 - \frac{l(l+1)}{2} (b^2 -b -2) -2 \right\rbrace \psi
- \left\lbrace 4b^3 \ddot{b} + 6b^2 \dot{b}^2 + l(l+1) b^2 \right\rbrace \eta
\right. \\ \left.
+ b^2 \dot{b} \left(b + 1 \right) \dot{\chi}
+ b \dot{b} \left( b^2 + b + \frac{1}{2} \right) \dot{\psi}
- 2b^3 \dot{b} \dot{\eta} -  2b^2 \dot{b} \zeta' + b^2 \ddot{\psi} - b \psi''
\right]  =0,
\end{split}
\end{equation}
and
\begin{equation}\label{29g}
\begin{split}
-\frac{1}{4\rho_0 b^6} \left[ \left\lbrace  6b^5 \ddot{b} - 2b^4 \ddot{b} - 4b^3 \dot{b}^2 + 9b^4 \dot{b}^2 + l(l+1)(3b^3 - b^4) \right\rbrace \chi
\right. \\ \left.
+  \left\lbrace 6b^5 \ddot{b} - 2b^4 \ddot{b} - 4b^3 \ddot{b}
+ 9b^4 \dot{b}^2 - 4b^3 \dot{b}^2 - 10b^2 \dot{b}^2 - 4b^2 +  l(l+1) (3b^3 + 2b^2 - b^4) \right\rbrace \psi
\right. \\ \left.
- \left\lbrace 16 b^4 \dot{b}^2 + 2b^4 \dot{b}^2 + 12b^5 \ddot{b} - 2 l(l+1)b^4
\right\rbrace \eta
+ \left\lbrace 3b^5 \dot{b} + 4b^4 \dot{b} \right\rbrace \dot{\chi} + \left\lbrace 3b^5 \dot{b} + 4b^4 \dot{b} + 3b^3 \dot{b} \right\rbrace \dot{\psi}
\right. \\ \left.
- 6b^5 \dot{b} \dot{\eta} - 12b^4 \dot{b}  \zeta' + 2b^5 \ddot{\chi} + \left\lbrace 2b^5 + 2b^4 \right\rbrace \ddot{\psi}
+ 2b^5 \chi'' + (2b^5 - 6b^3) \psi'' - 4b^5 \eta'' - 4b^5 \dot{\zeta}'
 \right]  =0.
\end{split}
\end{equation}

\vspace{0.5cm}
The evolution of $ w $, $ v $ and $ \Delta $ is governed by the following set of equations :
\begin{equation}\label{29h}
\begin{split}
\dot{w} = \frac{1}{8 b^{5/2} (2\dot{b}^2 +1)} \left[ -5 l(l+1) b \dot{b} \zeta - 2 l(l+1) b^2 \dot{\zeta} + (4b^3 \ddot{b} + 14b^2\dot{b}^2) \chi'
 + (4b^3 \ddot{b} - 6b \ddot{b} + 14b^2 \dot{b}^2 - 9\dot{b}^2) \psi'
\right. \\ \left.
 - (8b^3 \ddot{b} + 28b^2 \dot{b}^2) \eta' + 4b^3 \dot{b} \dot{\chi}'
 + (4b^3 \dot{b} + 4b \dot{b}) \dot{\psi}' -8b^3 \dot{b} \dot{\eta}' +4b^2 \ddot{\psi}' \right],
\end{split}
\end{equation}

\begin{equation}\label{29i}
\begin{split}
\dot{v} = -\frac{1}{8 b^{5/2} (2\dot{b}^2 +1)} \left[ (b^{5/2} \ddot{b} - 3b^{7/2} \ddot{b} + 2b^{2} \ddot{b} + 3b^{3/2} \dot{b}^2 + 5b \dot{b}^2
- 12b^{5/2} \dot{b}^2) \chi
+ (2b^{2} \ddot{b} + b^{5/2} \ddot{b} + 4b^{3/2} \ddot{b}
 \right. \\ \left.
 - 3b^{7/2} \ddot{b} + 8b^{1/2} \dot{b}^2 + 5b \dot{b}^2 - 12b^{5/2} \dot{b}^2 + 3b^{3/2} \dot{b}^2) \psi + (6b^{7/2} \ddot{b} + 24b^{5/2} \dot{b}^2) \eta
+ (2b^{2} \dot{b} - 7b^{5/2} \dot{b} - 3b^{7/2} \dot{b}) \dot{\chi}
\right. \\ \left.
 + (-2b^{3/2} \dot{b} - 3b^{7/2} \dot{b} - 7b^{5/2} \dot{b} + 2b^{2} \dot{b}) \dot{\psi} + 6b^{7/2} \dot{b} \dot{\eta}
 + 8b^{5/2} \dot{b} \zeta' - 2b^{7/2} \ddot{\chi} - (2b^{5/2} + 2b^{7/2}) \ddot{\psi} + 2b^{7/2} \dot{\zeta}' \right],
\end{split}
\end{equation}

\begin{equation}\label{29j}
\begin{split}
\dot{\Delta} = -\frac{1}{b^{3} (2\dot{b}^2 +1)} \left[ \left\lbrace 4b^3 \dot{b} \ddot{b} - 2b^2 \dot{b} \ddot{b} + 6b^2 \dot{b}^3 - 2b \dot{b}^3 + l(l+1)b \dot{b} \right\rbrace \chi
+ \left\lbrace 4b^3 \dot{b} \ddot{b} - 2b^2 \dot{b} \ddot{b} - 6b \dot{b} \ddot{b} + 6b^2 \dot{b}^3 - 2b \dot{b}^3 - 3\dot{b}^3 - \dot{b}
\right. \right. \\ \left. \left.
 + l(l+1) \dot{b} \left(b + \frac{1}{2}\right) \right\rbrace \psi  - \left\lbrace 8b^3 \dot{b} \ddot{b} + 12b^2 \dot{b}^3 \right\rbrace \eta
+ \left\lbrace b^3 \ddot{b} + 2b^3 \dot{b}^2 + 2b^2 \dot{b}^2 + \frac{l(l+1)}{2} b^2 \right\rbrace \dot{\chi}
+ \left\lbrace b^3 \ddot{b} + \frac{3}{2}b^2 \ddot{b} + 2b^3 \dot{b}^2 + 2b^2 \dot{b}^2
\right. \right. \\ \left. \left.
+ \frac{l(l+1)}{2} (b^2 +b) - b \right\rbrace  \dot{\psi}
- 4b^3 \dot{b}^2 \dot{\eta} - (2b^3 \ddot{b} + 6b^2 \dot{b}^2) \zeta'
+ b^3 \dot{b} \ddot{\chi} + (b^3 \dot{b} + \frac{3}{2}b^2 \dot{b}) \ddot{\psi} - 2b \dot{b} \psi'' - 2b^3 \dot{b} \dot{z}' - b^2 \dot{\psi}''  \right].
\end{split}
\end{equation}
These equations are obtained by taking time-derivatives of equations \eqref{29a}, \eqref{29b} and \eqref{29e}, and using the continuity equation \eqref{E2}. Following \cite{CL1}, we can say that the polar perturbation equations \eqref{29b}, \eqref{29d}, \eqref{29f}, \eqref{29g} and \eqref{29i} hold for $ l\geq 1 $, and equations \eqref{29a}, \eqref{29e}, \eqref{29h} and \eqref{29j} hold for $ l \geq 0 $.

\section*{Acknowledgement}
The authors are thankful to the anonymous reviewers for the useful comments to improve the quality of the paper. SD acknowledges the financial support from INSPIRE (AORC), DST, Govt. of India (IF180008). SG thanks IUCAA, India for an associateship and CSIR, Government of India for the major research grant [No. 03(1446)/18/EMR-II].

\end{document}